\documentclass[conference]{IEEEtran}
\makeatletter
\def\ps@headings{%
\def\@oddhead{\mbox{}\scriptsize\rightmark \hfil \thepage}%
\def\@evenhead{\scriptsize\thepage \hfil \leftmark\mbox{}}%
\def\@oddfoot{}%
\def\@evenfoot{}}
\makeatother
\usepackage{amsmath}
\usepackage{amsfonts}
\usepackage{bm}
\usepackage{graphicx}
\usepackage{algorithmic}
\usepackage{algorithm}
\usepackage{color}
\usepackage{url}
\usepackage{verbatim}

\newtheorem{prop}{Proposition}

\begin{document}
\title{Dual-Level Compressed Aggregation: \\ Recovering Fields of Physical Quantities from \\ Incomplete Sensory Data}

\author{
\IEEEauthorblockN{Liu Xiang\IEEEauthorrefmark{1},
Jun Luo\IEEEauthorrefmark{1},
Chenwei Deng\IEEEauthorrefmark{1},
Athanasios V. Vasilakos\IEEEauthorrefmark{2} and
Weisi Lin\IEEEauthorrefmark{1}}
\IEEEauthorblockA{\IEEEauthorrefmark{1}School of Computer Engineering, Nanyang Technological University, Singapore}
\IEEEauthorblockA{\IEEEauthorrefmark{2}Department of Electrical and Computer Engineering, National Technical University of Athens, Greece\\
Email: \{xi0001iu, junluo, cwdeng\}@ntu.edu.sg, vasilako@ath.forthnet.gr, wslin@ntu.edu.sg}
}

\maketitle

\begin{abstract}
Although \textit{wireless sensor networks} (WSNs) are powerful in monitoring physical events, the data collected from a WSN are almost always \textbf{incomplete} if the surveyed physical event spreads over a wide area. The reason for this incompleteness is twofold: i) insufficient network coverage and ii) data aggregation for energy saving. Whereas the existing recovery schemes only tackle the second aspect, we develop \textit{Dual-lEvel Compressed Aggregation} (DECA) as a novel framework to address both aspects. Specifically, DECA allows a high fidelity recovery of a widespread event, under the situations that the WSN only sparsely covers the event area and that an in-network data aggregation is applied for traffic reduction. Exploiting both the low-rank nature of real-world events and the redundancy in sensory data, DECA combines \textit{matrix completion} with a fine-tuned \textit{compressed sensing} technique to conduct a dual-level reconstruction process. We demonstrate that DECA can recover a widespread event with less than 5\% of the data, with respect to the dimension of the event, being collected. Performance evaluation based on both synthetic and real data sets confirms the recovery fidelity and energy efficiency of our DECA framework.
\end{abstract}


\section{Introduction} \label{sec:intro}
\begin{figure}[ht]
    \parbox{\columnwidth}{\parbox{.49\columnwidth}{\center\includegraphics[width=.49\columnwidth]{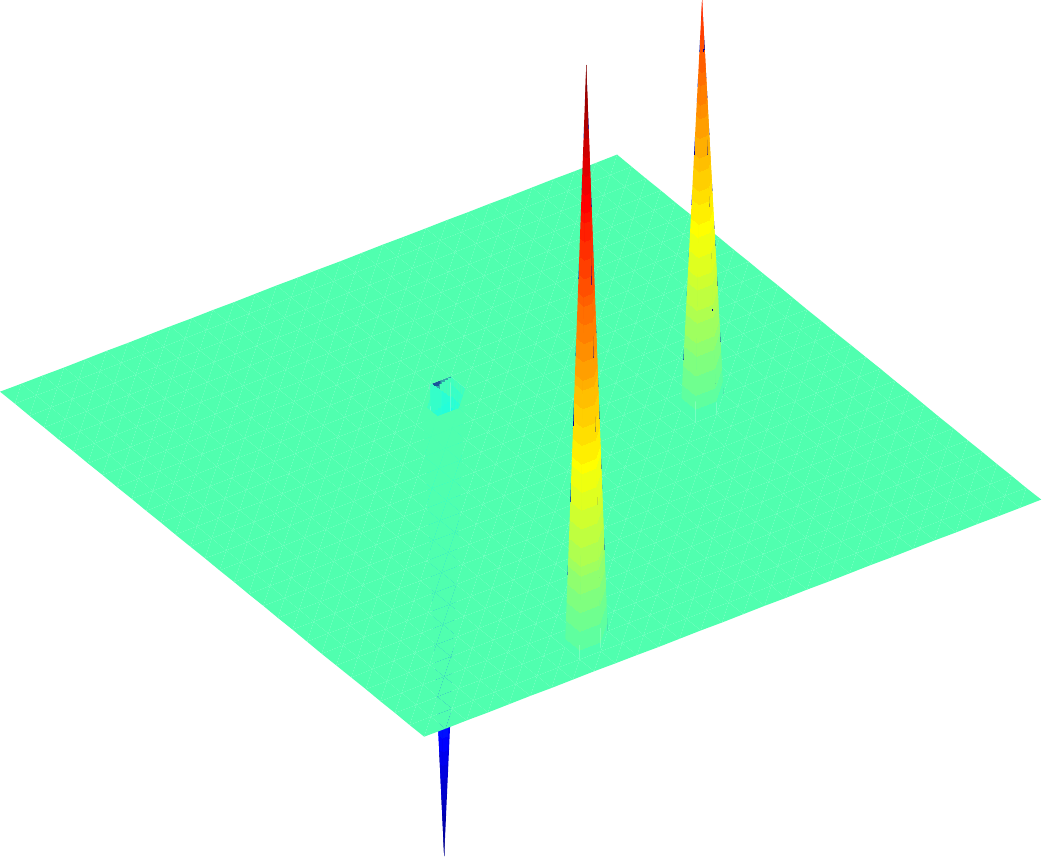}}
                              \parbox{.49\columnwidth}{\center\includegraphics[width=.49\columnwidth]{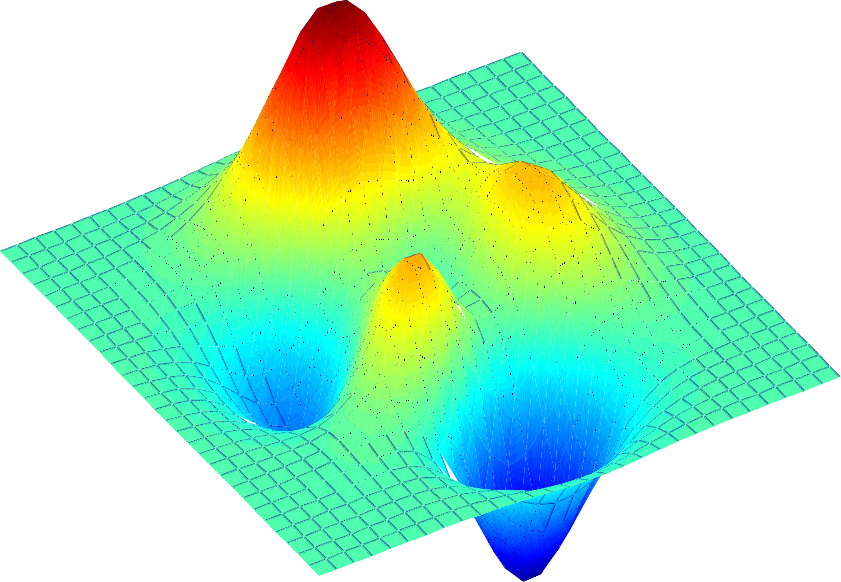}}}
    \parbox{\columnwidth}{\parbox{.49\columnwidth}{\center\scriptsize(a)~Burst events}
                              \parbox{.49\columnwidth}{\center\scriptsize(b)~Field}}
    \caption{Two types of physical events.} \label{fig:phyevents}
  \end{figure}

  As \textit{wireless sensor networks} (WSNs), with their networked sensors, have the ability of ``merging"  into physical environments, they are generally considered as powerful tools to survey or monitor physical events. Several real systems have been emerged, including FireFly~\cite{Firefly} that tracks the position of miners, GreenOrbs~\cite{greenorbs} that collects ecological information from a forest, and many others. We may roughly categorize the events subject to WSNs' surveillance into two types. On one hand, \textit{burst events} take place only sporadically, and monitoring such events often boils down to detecting abnormal changes in an area. For example, a sudden temperature change in a warehouse may signal a fire alarm. On the other hand, a \textit{field} (of certain physical quantities, e.g., humidity or pollution level) has a smooth distribution over a wide area and usually undergoes gradual changes. We illustrate these two types of events in Fig.~\ref{fig:phyevents}.

  Whereas monitoring burst events may only require intermittent data transmissions across a WSN to report abrupt changes, surveying a field does demand a constant data gathering from a large-scale WSN that is meant to sufficiently cover the monitored area. Obviously, in providing high fidelity field surveillance, the energy efficiency issue of WSNs becomes a bottleneck. In this paper, we aim at tackling the conflict between sensory data fidelity and energy efficient data gathering for WSNs that perform field surveillance.

  \subsection{Problem Overview and Motivations} \label{sec:overmotiv}
    Energy efficient data gathering in WSNs has been a long standing topic since the inception of these networked sensing systems. While the approaches involving routing or scheduling focus on improving the efficiency of data transportation \cite{GanesanCBL-ToSN06,FanLS-TMC07,WuLLL-TPDS10}, \textit{data aggregation}\footnote{We use the term ``data aggregation'' in a broad sense: it refers to any in-network traffic reduction mechanism.} directly fights the redundancy in sensory data, striving to significantly reduce the amount of data to be transported (e.g., \cite{Madden-OSR02,Cristescu-ToN06,Gupta-ToSN08}). Consequently, data aggregation is often deemed as a crucial mechanism to achieve energy efficient data gathering for WSNs.

    \subsubsection{Conventional Data Aggregation}
      In general, data aggregation can be either lossy or lossless. \textit{Lossy aggregation} usually adopts simple aggregation functions (e.g., MIN, MAX, or SUM) and only extracts certain features from the sensory data (e.g., \cite{Madden-OSR02}). Obviously, though this approach may improve the efficiency of monitoring burst events, it is definitely not suitable for field surveillance, as, apart from a few features, most of the information about a field is lost and is beyond recoverability. \textit{Lossless aggregation}\footnote{Lossless aggregation based on lossy compression techniques (e.g., wavelet compression) may still sacrifice data fidelity. Therefore, lossless aggregation is so termed to emphasize its intention to preserve the field information, rather than only extracting a few features.} is closely related to data compression: they both aim at ``squeezing'' the redundancy or insignificant components of a given data set to reduce its volume \cite{Cristescu-ToN06,Gupta-ToSN08}. However, unlike common data compression (where the underlying statistic model is known \textit{\`{a} priori} or can be easily discovered), the model that describes data correlations for a sensing field is often unknown or may vary in time. As a result, distributed source coding techniques \cite{Cristescu-ToN06} using, for example, Slepian-Wolf coding are not exactly practical. Moreover, collaboratively discovering the data correlation structure \cite{Gupta-ToSN08} leads to high communication load that offsets the benefit of this aggregation technique.

      If we consider the field under WSN surveillance as an ``image'', image compression techniques (e.g., DCT or wavelet based) appear to be a good way to realize lossless aggregation \cite{Ciancio-IPSN06}. Unfortunately, this approach is facing several major difficulties. Firstly, unless sensors are deployed to monitor every ``pixel'' of a field, the sensory data are not amenable to a 2D transformation. However, WSNs can be randomly deployed in order to avoid high cost at the deployment phase. Secondly, even if 2D transformations can be applied to a regularly deployed WSN, applying such transformations in-network can bring high overhead, due to the need of exchanging coefficients among coding nodes. Finally, given the difficulty in using 2D transformations, taking into account higher dimensional correlation of the field data (e.g., the temporal correlation) becomes almost impossible.

    \subsubsection{Compressed Sensing Based Data Aggregation} \label{sec:cda}
      Following several celebrated works in signal processing~\cite{CandesRT-TIT06, Donoho-TIT06}, \textit{compressed sensing} (CS) -- the technique for finding sparse solutions to underdetermined linear systems -- has been intensively studied. It suggests an easy way to acquire and reconstruct a signal given that it is sparse or compressible. Right after its development, CS was introduced into WSNs as a data aggregation technique \cite{Haupt-SPMag08,QuerMMRWZ-ITA09,LeePSKO-GSN09,LuoWSC-MobiCom09,LuoXR-ICC10,LeeO-ASC10}. CS promises to deliver, with high probability, a full recovery of signals from far fewer samples than their original dimension, as long as the signals are sparse or compressible in some domain \cite{Candes-SPMag08}. In fact, the encoding process does not rely on the data correlation structure and the sensor nodes are not supposed to be aware of the correlation~\cite{Haupt-SPMag08}, which directly translates to the model-less ``compression'' and the blind encoding. In addition, the in-network aggregation required by CS incurs very light computation load \cite{LuoXR-ICC10}. All these make CS aggregation very attractive.

      However, three main issues are still hampering the practical use of CS aggregation. Firstly, as the existing techniques make use of conventional signal/image compression domains (e.g., DCT domain) for sensory data, the need for regular WSN deployments persists \cite{QuerMMRWZ-ITA09}. Secondly, even if one could fully recover the sensory data obtained by a regularly deployed WSN, there is no guarantee that these data can faithfully represent the field under surveillance, as the size of a WSN is often insufficient to cover a field. Last but not least, designing energy efficient routing is highly nontrivial as it may involve the coherence between the sparse domain and the network topology \cite{LeePSKO-GSN09,LuoXR-ICC10,LeeO-ASC10}.

  \subsection{Our Approach and Contributions}
    We first acknowledge that it is an ill-posed problem to directly recover a surveyed field from CS aggregated (hence under-sampled) data. Our response is the \textit{Dual-lEvel Compressed Aggregation} (DECA) framework. In essence, DECA recovers, at the first level, the sensory data obtained by the whole WSN from the CS aggregated data. Then at the second level, DECA recovers the field based on the outcome of the first level. This decomposition in CS recovery brings several great benefits, in response to the hampering issues we mentioned above. First of all, adequate CS techniques can be applied to individual levels to achieve the best recovery performance and to avoid the requirement for regular WSN deployments. Secondly, the field can be recovered from the CS aggregated data, even though the original sensory data are only random samples of the field. Finally, an energy efficient routing technique can be deployed without incurring too much complexity for in-network coordinations.

    In proposing our DECA framework, we are making the following main contributions:
    \begin{itemize}
      \item We propose, for the first time, the concept of dual-level CS aggregation and field recovery, dedicated to WSNs that monitor smoothly distributed (both spatial and temporal) physical quantities.
      \item We apply \textit{diffusion wavelets} to the first-level recovery (from CS aggregated data to the original sensory data), and we propose novel diffusion operators to achieve the best recovery performance. These operators also allow temporal correlation to be naturally taken into account.
      \item We apply \textit{matrix completion} scheme to the second-level recovery (from sensory data to fields). We discover that the performance is as good as if the sensory data were directly collected, although they are actually recovered from CS aggregated data.
      \item We show that, under the DECA framework, the in-network computation is extremely light for sensor nodes, and natural tree partitions of a WSN can lead to a significant energy saving for the overall data collection process.
    \end{itemize}

  \subsection{Roadmap}
    In the remaining of our paper, we first survey the related literature on CS aggregation in Sec.~\ref{sec:related}. Then we present the basic principles concerning the building blocks of DECA in Sec.~\ref{sec:bgr}. We focus on the design of DECA in Sec.~\ref{sec:deca}. We evaluate the performance of DECA in terms of both data fidelity and energy efficiency in Sec.~\ref{sec:perf}. Finally, we conclude our paper in Sec.~\ref{sec:con}.

\section{Related Work} \label{sec:related}
  Our discussions in this section only emphasize on CS aggregation and related applications of CS to networking issues. As we explained in Sec.~\ref{sec:overmotiv}, a lossy aggregation is meaningful only to burst events, while a lossless aggregation either requires model awareness or regular sensor deployments. Given the absence of parallels between DECA and these approaches (as DECA demands none of these prerequisites), we omit their discussions.

  \subsection{CS Data Aggregation in WSNs}
    Two of the earliest contributions in applying CS to WSN data collection are by Bajwa \textit{et al.}\ \cite{BajwaHSN-IPSN06} and Duarte \textit{et al.}\ \cite{DuarteWBB-IPSN06}. However, while \cite{BajwaHSN-IPSN06} only involves single-hop communications and is hence not really concerning data aggregation, \cite{DuarteWBB-IPSN06} focuses on compressing temporally correlated data, while relying on existing protocols to take care of multi-hop communications.

    Quer \textit{et al.}\ \cite{QuerMMRWZ-ITA09} investigated the CS aggregation performance along with routing costs in multi-hop WSNs. They concluded that the accuracy of CS recovery depends on routing paths. However, this is an artifact introduced by defining the \textit{sensing matrix} (see Sec.~\ref{sec:cs}) of CS according to the routing paths. Moreover, as we mentioned in Sec.~\ref{sec:cda}, \cite{QuerMMRWZ-ITA09} requires a regular WSN deployment on a grid of cells and a full coverage of the surveyed field, i.e., one sensor per cell.

    Lee \textit{et al.}\ \cite{LeePSKO-GSN09} also targeted the CS aggregation issue. They aimed at identifying proper network partitions for energy efficient CS aggregation. The main conclusion drawn in \cite{LeePSKO-GSN09} is that, if CS aggregations are performed for individual partitions of a WSN, the sensing matrix has to take the characteristics of the sparse domain into account. This, unfortunately, contradicts the spirit of CS, i.e., sensing matrices can be random. As we will show in Sec.~\ref{sec:routing}, if a sparse domain is properly chosen, the signal energy most concentrates on the ``low frequency'' components. Therefore, simple sensing matrices (e.g., Bernoulli random matrix) still suffice for CS aggregation performed for individual partitions.

    Two later (independent) proposals \cite{LuoWSC-MobiCom09} and \cite{LuoXR-ICC10} gave more emphasis on routing efficiency in CS aggregation. \cite{LuoWSC-MobiCom09} proved that, if $k$ random samples are aggregated from a WSN of $n$ nodes, the throughput is $\frac{n}{k}$ times higher. However, a more detailed investigation (involving an interference model and scheduling) in \cite{LuoXR-ICC10} showed that the \textit{plain CS aggregation} used in \cite{LuoWSC-MobiCom09} may have a throughput even lower than no aggregation at all (\textit{non-aggregation} hereafter). \cite{LuoXR-ICC10} further proposed the so called \textit{hybrid CS aggregation}; it achieves a throughput always better than non-aggregation. \cite{LuoWSC-MobiCom09} also evaluated the performance of CS recovery. However, only data sampled from one-dimension signals (or sampled in 2D but can be reduced to vectors) are treated.

  \subsection{Other Applications of CS in Networking}
    Applications of CS to other networking issues came earlier than CS aggregation in WSNs. These applications are mostly concerned with traffic measurements in the Internet. Coates \textit{et al.}\ \cite{CoatesPR-IMC07} exploited the performance correlations between overlapping paths and proposed to use CS to reduce the number of measurements. Their proposal of using diffusion wavelets to accommodate measures taken on an arbitrary network topology has motivated our first-level recovery in DECA. However, the design of diffusion operators are totally different due to distinct correlation structures in data.

    Leveraging on the low rank feature of the Internet \textit{traffic matrices} (TMs), Zhang \textit{et al.}\ \cite{ZhangRWQ-SIGCOMM09} applied the \textit{matrix completion} technique (the most up-to-date development in CS) to recover TMs from highly incomplete samples. They demonstrated that the matrix completion technique consistently outperforms other commonly used methods such as \textit{singular value decomposition} (SVD). We make a different use of matrix completion in the second-level recovery of DECA, considering the monitored field as an ``image'' and hence a matrix. The low rankness of this matrix is obvious: as far as an image does not just contain noise, it always has a sparse or compressible representation under  SVD.

    Applications of CS in wireless networks other than data aggregation also appeared recently. Charbiwala \textit{et al.}\ \cite{CharbiwalaCZKSHB-Infocom10} proposed to use CS as a kind of erasure coding strategy for forward error correction, aiming at improving the robustness for data transmission. Rallapalli \textit{et al.}\ \cite{RallapalliQZC-MobiCom10} looked at the localization problem in mobile networks. The rationale for applying CS techniques is twofold: the matrix of node coordinates can be well approximated by a low-rank matrix, and the mobility features of nodes are temporally correlated.

\section{Background and Rationales} \label{sec:bgr}
  In this section, we introduce the principles of DECA's building blocks. The first three topics, namely compressed sensing, diffusion wavelets, and matrix completion, are concerning data sampling and recovery procedures, and the last topic deals with efficient routing structure to support CS compression.

  \subsection{Compressed Sensing (CS)} \label{sec:cs}
    The theory of CS is pioneered by Cand\`{e}s and Tao~\cite{CandesRT-TIT06}, as well as Donoho~\cite{Donoho-TIT06}, and later developed by many others (e.g., \cite{Haupt-SPMag08}). The theory asserts that one can recover a certain data set from far fewer samples, as long as the data set has a sparse representation in a domain, and the sampling process is largely incoherent with the basis that enables the sparse representation.

    Suppose an \textit{n-dimensional} vector $\mathbf{u}$ is \textit{m-sparse} under a proper domain spanned by $\Psi=[\boldsymbol{\psi}_1,\dots,\boldsymbol{\psi}_n]$, where $\boldsymbol{\psi}_i$ represents a column vector of the basis, we have
    \begin{eqnarray}
      \mathbf{u} = \Psi \mathbf{w} = \sum_{i=1}^m w_i \boldsymbol{\psi}_i,~~~~~~~\mbox{for}~~m \ll n,
    \end{eqnarray}
    where $\mathbf{w}$ is called the \textit{sparse representation} of $\mathbf{u}$: it has only $m \ll n$ non-zero entries. Then the CS theory suggests that, under certain conditions, instead of directly computing and collecting the compressed coefficients $\mathbf{w}$, we may collect a slightly longer vector $\mathbf{v} = \Phi\mathbf{u}$, where $\Phi=[\boldsymbol{\phi}_1,\dots,\boldsymbol{\phi}_n]$ is a $k \times n$ ``sensing" matrix corresponding to the sampling process. Consequently, we can recover $\mathbf{u}$ from $\mathbf{v}$ with high probability by solving the following convex optimization problem
    \begin{eqnarray}
      \underset{\mathbf{w} \in \mathbb{R}^n}{\mathrm{minimize}} && \|\mathbf{w}\|_{\ell_1} = \sum_i |w_i| \label{eq:cs} \\
      \mbox{subject to} && ~~~~\mathbf{v} = \Phi \Psi \mathbf{w}, \nonumber
    \end{eqnarray}
    and by letting $\mathbf{u} = \Psi \mathbf{\hat{w}}$, with $\mathbf{\hat{w}}$ being the optimal solution of~(\ref{eq:cs}). We hereafter refer to the random sampling process $\mathbf{v} = \Phi\mathbf{u}$ as \textit{CS coding}, and the process of recovering $\mathbf{u}$ by solving~(\ref{eq:cs}) as \textit{CS recovery}.

    The \textit{incoherence} \cite{CandesRT-TIT06} between the sensing matrix $\Phi$ and the sparse basis $\Psi$ is crucial to the recovery performance. In practice, a $\Phi$ with Gaussian or Bernoulli entries largely abides by the incoherence condition for any $\Psi$ if the number of measurements satisfies $k \ge \mathcal{O}(m\log n)$ \cite{CandesRT-TIT06}. The choices for $\Psi$ include Fourier basis, wavelet basis, DCT basis, etc., depending on the specific applications. What really makes CS attractive is that the sparse basis does not need to be known during the encoding process, which makes it extremely suitable for data aggregation in WSNs.

    If the vector $\mathbf{u}$ is not exactly sparse but only compressible or if the sampling comes with errors, the constraint in problem~(\ref{eq:cs}) needs to be replaced by
    \begin{eqnarray}
      \|\Phi \Psi \mathbf{w} - \mathbf{v} \|_{\ell_2} &\le& \epsilon,
    \end{eqnarray}
    Now, the theory requires $\Phi\Psi$ to obey the so-called \textit{restricted isometry principle} (RIP) in order to guarantee a successful recovery~\cite{Candes-SPMag08}. In practice, for any fixed $\Psi$, RIP holds with high probability if $\Phi$ has i.i.d.\ entries from the normal distribution $\phi_{i,j}\sim N(0,1/k)$ or from a symmetric Bernoulli distribution $\mathrm{Pr}(\phi_{i,j}=\pm 1/\sqrt{k})=1/2$, and if $k \ge \mathcal{O}\left(m\log(n/m)\right)$.

  \subsection{Diffusion Wavelets} \label{sec:diffwave}
    Although CS allows a flexible choice for sparse bases, most of the sparse bases work only for vectors sampled from 1D signal.\footnote{Even for 2D signals such as images, they are sampled in a 1D manner (row by row or column by column) to adapt to sparsifying transformations (e.g., wavelet transform).} In order to cope with vectors sampled on manifolds or graphs (e.g., data sensed by a WSN), \textit{diffusion wavelets} are developed to generalize classic wavelets \cite{CoifmanM-04}. As opposed to dilating a ``mother wavelet'' by powers of two to generate a set of classic wavelet bases, the dyadic dilation that generates diffusion wavelets relies on a \textit{diffusion operator}. Here diffusion is used as a smoothing and scaling tool to enable multiscale analysis on manifolds or graphs.

    Let us take an arbitrary graph $G$ as an example to illustrate the idea. Suppose the weighted adjacency matrix of $G$ is $\Omega=[\omega_{i,j}]$, where $\omega_{i,j}$ is the weight of edge $(i,j)$. Let $\Lambda=[\lambda_{i,j}]$ be the \textit{normalized Laplacian} of $G$, the definition is given below:
    \begin{eqnarray}
    \lambda_{i,j} =
      \begin{cases}
      1 & i=j, \\
      - \frac{\omega_{i,j}}{\sqrt{\sum_{p}{\omega_{i,p}} \sum_{p}{\omega_{p,j}}}} & \mathrm{otherwise}.
      \end{cases}
    \label{eq:lap}
    \end{eqnarray}
    It is well known that $\Lambda$ characterizes the degree of correlations between function values taken at vertices of the graph $G$ \cite{Chung97}. Roughly speaking, each eigenvalue (and the corresponding eigenvector) represents the correlation under a certain scale. In order to decompose the signal sampled on a graph in a multiscale manner, one may consider partitioning the range space of $\Lambda$. The idea behind diffusion wavelets is to construct a \textit{diffusion operator} $O$ from $\Lambda$, such that they share the same eigenvectors whereas all eigenvalues of $O$ are smaller than 1. Consequently, recursively raising $O$ to power 2 and applying a fixed threshold to remove the diminishing eigenvalues (hence the corresponding eigenvectors and the subspaces spanned by them) lead to a  dilation of the null space but a shrinkage of the range space; this naturally produces space splitting.

    More specifically, $O^{2^j}$ is computed at the $j$-th scale, eigenvalue decomposition is derived for it, and the resulting eigenvectors form a basis that (qualitatively) represents the correlation over neighborhood of radius $2^j$ hops on the graph. Denote the original range space of $O$ by $U_0 = \mathbb{R}^n$, it is split recursively: at the $j$-th level, $U_{j-1}$ is split into two orthogonal subspaces: the scaling subspace $U_{j}$ that is the range space of $O^{2^j}$, and the wavelet subspace $V_{j}$ as the difference between $U_j$ and $U_{j-1}$. Given a specified decomposition level $\gamma$, the diffusion wavelet basis $\Psi$ is the concatenation of the orthonormal bases of $V_{1}, \dots, V_{\gamma}$ and $U_{\gamma}$. Interested readers are referred to~\cite{CoifmanM-04} for detailed exposition. We want to point out that different diffusion operators lead to different wavelet bases, therefore the art of our later design lies in the proper choice of an operator.

  \subsection{Matrix Completion (MC)} \label{sec:mc}
    As an extension to the classic CS techniques that work in vector spaces, similar results have been developed for matrices, under the name of \textit{matrix completion} (MC) \cite{CandesP-ProcIEEE10}. The assertion is similar: a matrix can be recovered from a small set of samples of its entries, as far as it is low rank and its singular vectors are reasonably spread across all coordinates.

    Given a low rank matrix $M \in \mathbb{R}^{n_1 \times n_2}$ with rank $\zeta$ and an observation of $k$ entries of $M$ through a sampling operator $\mathcal{P}_{\Pi}(M)$ such that
    \begin{eqnarray}
    [\mathcal{P}_{\Pi}(M)]_{ij} =
      \begin{cases}
      M_{ij} & (i,j) \in \Pi, |\Pi| = k \\
      0 & \mathrm{otherwise},
      \end{cases}
    \nonumber
    \end{eqnarray}
    where $\Pi$ is a subset of $M$'s entries. If the number of samples $k$ and the matrix $M$ satisfy certain conditions, we may recover with high probability the whole $M$ from the $k$ samples by solving the following \textit{nuclear-norm} minimization problem:
    \begin{eqnarray}
      \mathrm{minimize} && \|X\|_\ast ~= \!\!\sum_{i=1}^{\min(n_1,n_2)} \sigma_i \label{eq:mc} \\
      \mbox{subject to} && ~~~~\mathcal{P}_{\Pi}(X) = \mathcal{P}_{\Pi}(M). \nonumber
    \end{eqnarray}
    This typical convex optimization problem is amenable to efficient solution techniques.

    The condition in terms of $k$ is $k \ge \mathcal{O}(\max(n_1,n_2) \zeta \log^\tau\!(\max(n_1,n_2)))$, where $\tau$ is some constant \cite{CandesP-ProcIEEE10}. The conditions for $M$ may have various representations, and they generally confine the sparsity of the singular vectors of $M$ (or \textit{incoherence} of $M$). Intuitively speaking, if some singular vectors of $M$ are very sparse, random sampling may well miss those non-zero components and hence fail to preserve the structure of $M$. Different representations of the incoherence condition can lead to slight changes to the condition for $k$ (e.g., different constant $\tau$), and sometimes even a different recovering algorithm \cite{CandesP-ProcIEEE10}.

    We propose to use MC for recovering an image $M$ from incomplete pixel samples, which differs from its original intention. Compared with the classic CS technique (discussed in Sec.~\ref{sec:cs}), MC differs in two ways. Firstly, MC has no coding procedure, apart from a random sampling. Secondly, MC does not require a sparse domain to be explicitly chosen for the recovery, as the domain is suggested by SVD of the matrix. While the first feature allows a WSN (whose size is much smaller than the dimension of $M$) to be randomly deployed for monitoring $M$, the second feature makes the recovery independent of the targeted $M$.

  \subsection{Compressed Data Aggregation (CDA)} \label{sec:csrt}
\begin{figure}[htb]
      \centering
      \includegraphics[width=\columnwidth]{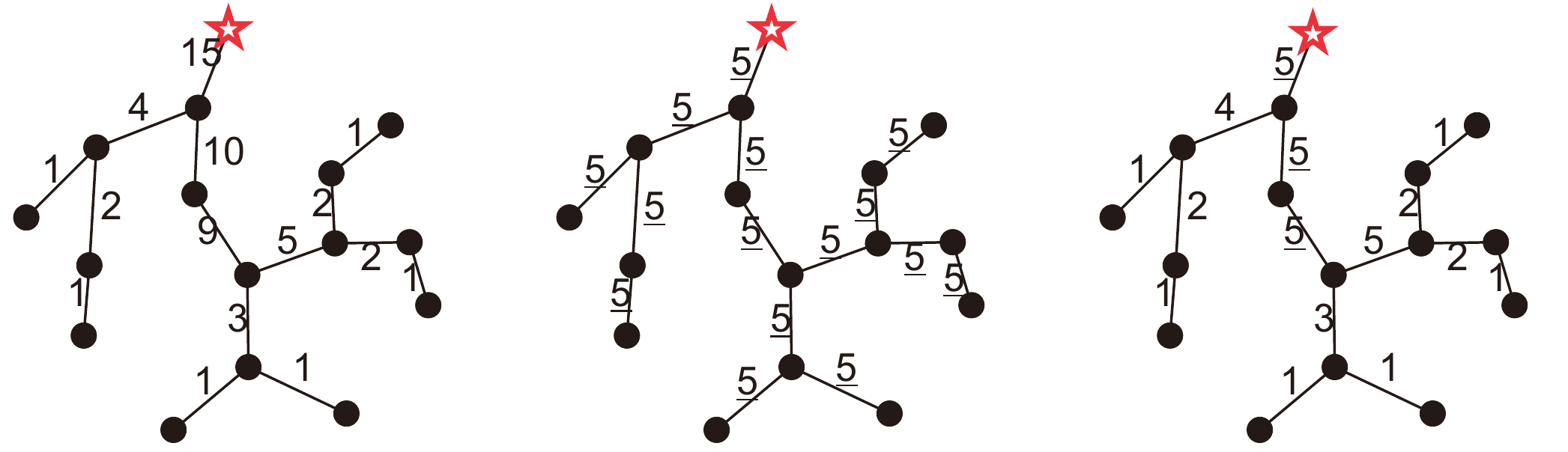}
      \parbox{.3\columnwidth}{\center\scriptsize(a) Non-aggregation}
      \parbox{.33\columnwidth}{\center\scriptsize(b) Plain CS aggregation}
      \parbox{.33\columnwidth}{\center\scriptsize(c) Hybrid CS aggregation}
      \caption{Different aggregations on a tree. The red stars denote the sink. Numbers beside links represent the traffic load with underlines indicating the CS coded traffic.}
      \label{fig:aggtrees}
    \end{figure}

    Assume we are given a WSN of $n$ nodes with each one acquiring a sample $u_i$, the sink is supposed to collect all data $\mathbf{u} = [u_1, \cdots, u_n]^T$. Without data aggregation, clearly, the most energy efficient routing strategy is a \textit{shortest path tree} (SPT) rooted at the sink, where the nodes around the sink tend to carry heavy traffic load, as shown in Fig.~\ref{fig:aggtrees}(a).

    The recent developed CS theory suggests a way to relieve the bottleneck \cite{Haupt-SPMag08, LuoWSC-MobiCom09}. Let us rewrite the random sampling in a column form $\mathbf{v} = u_1 \boldsymbol{\phi}_1 + \cdots + u_n \boldsymbol{\phi}_n$. For CS-based data aggregation, each node $i$ first ``expands" it own sensory data $u_i$ to $k$ coded items, which corresponds to $u_i\boldsymbol{\phi}_i$. These $k$ data items are then sent along a data collection tree. Whenever more than one set of such data items converge at a node, elements with the same indices are summed up. The eventual outcome accomplishes the overall CS coding $\mathbf{v} = \Phi\mathbf{u}$. The imposed identical flow (Fig.~\ref{fig:aggtrees}(b)), on one hand, eliminates the conventional bottleneck; but on the other hand, it introduces additional traffic to the leaf nodes.

    As an improvement, we proposed \textit{hybrid CS aggregation} in \cite{LuoXR-ICC10} that fully exploits the advantage of CS. In a nutshell, if the number of data items converged at a certain node is below $k$, no aggregation is performed. The CS aggregation starts to work only when $k$ or more data items gather at a node. The idea is illustrated in Fig.~\ref{fig:aggtrees}(c). Each aggregation is equivalent to a partial CS coding $\mathbf{v}' = \Phi'\mathbf{u}$, where $\Phi'$ contains a subset of all columns of $\Phi$.

    In our recent work \cite{XiangLV-SECON11}, we also investigated the energy efficient configurations for the hybrid CS aggregation through joint routing and CS aggregation. We have proven the \textit{minimum energy compressed data aggregation} problem is in general NP-complete by showing the equivalence between an optimal tree with $k=2$ and the \textit{maximum leaf spanning tree} (MLST) problem \cite{GareyJ79}. Then we designed an efficient greedy heuristic to obtain the near optimal configurations. The basic idea is to grow the ``core'' (the set of nodes that transmit $k$ samples) iteratively in a greedy manner until no node needs to be added to the core; the remaining nodes simply transmit non-aggregated data samples. We omit the algorithm details; interested readers are referred to~\cite{XiangLV-SECON11}.

    Our DECA framework makes use of the hybrid CS aggregation for joint CS coding and routing. However, instead of only considering a single tree, we partition a WSN into several trees and treat each tree independently, as will be addressed in Sec.~\ref{sec:routing}.

\section{DECA: Decomposed CS Aggregation and Field Recovery} \label{sec:deca}
  In this section, we first introduce our network model and formally define our problem in Sec.~\ref{sec:netmodel}, then we will present the two recovery levels and the aggregation mechanism for DECA in Sec.~\ref{sec:1level} to Sec.~\ref{sec:routing}.

  \subsection{Network Model and Problem Definition} \label{sec:netmodel}
    We assume a WSN is deployed to monitor a 2D area. Without loss of generality, we assume this 2D area has a rectangular shape. We partition this area into an $a \times b$ grid of square cells; the size of a cell represents the \textit{sensing coverage} of a node. Sensor nodes are randomly deployed with a \textit{coverage ratio} $\rho$, i.e., a cell is covered by a node with probability $\rho$. We represent the WSN by a connected graph $G(V, E)$, where the vertex set $V$ corresponds to the nodes in the network, and the edge set $E$ corresponds to the wireless links between nodes. One special node $s \in V$ is known as the sink; it collects data from the whole network. We denote by $n$ the cardinality of $V$. Obviously, we have $n = \rho(a\times b)$. Let $\mathbf{c}\!: E \rightarrow \mathbb{R}_{0}^{+}$ be a cost assignment on $E$, with $c(i,j)\!: (i,j) \in E$ being the energy expense of sending one unit of data across link $(i,j)$.

\begin{figure}[h]
      \centering
      \includegraphics[width=.8\columnwidth]{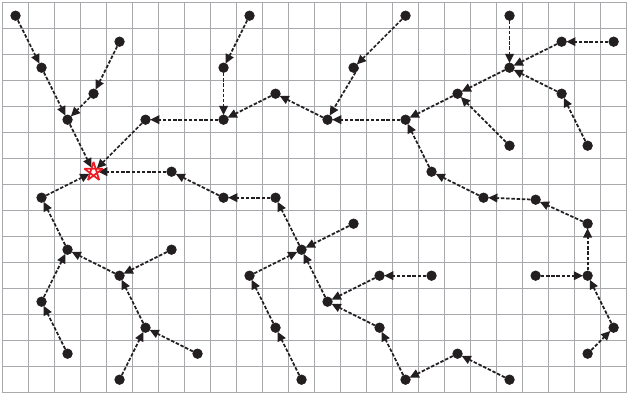}
      \caption{A WSN that monitors a rectangular area. The area is partitioned into a grid of square cells, the pentagram indicates the sink, and only the links used by the data collection tree are shown.} \label{fig:wsn}
    \end{figure}

    We assume that all nodes are roughly time synchronized and the data collection proceeds in rounds. At the beginning of a round $r$, node $i$ produces one sample $u^r_i$, and the sink collects all information at the end. In order to avoid duplicated aggregations on the way towards the sink, we restrict the data aggregation on a \textbf{tree} rooted at the sink. Finally, we assume that the sink knows the locations of all nodes. We illustrate our assumptions in Fig.~\ref{fig:wsn}.

    During a certain time period represented by a finite index set $\mathcal{R}$ for all the rounds within this period, the WSN produces a set of sensory data vectors $\{\mathbf{u}^r\}_{r \in \mathcal{R}}$, where $\mathbf{u}^r = [u^r_1, u^r_2, \cdots, u^r_n]^T$ is the data vector produced during round $r$. During the same period, the field under surveillance is represented by $\{F^r\}_{r \in \mathcal{R}}$ with $F^r$ being an $a \times b$ matrix that models the area at round $r$. Then $F^r_{ij}$ refers to the value of the monitored physical quantity at the $i$-th row and $j$-th column of the discretized area during round $r$. In order to reduce the energy consumption of the WSN, an in-network data aggregation is performed such that the sink collects $\{\mathbf{v}^r\}_{r \in \mathcal{R}}$, which is a compressed version of $\{\mathbf{u}^r\}_{r \in \mathcal{R}}$. Now our questions are the following:
    \begin{itemize}
      \item[\textbf{Q1:}] How can we recover $\{F^r\}_{r \in \mathcal{R}}$ from $\{\mathbf{v}^r\}_{r \in \mathcal{R}}$?
      \item[\textbf{Q2:}] What is the tradeoff relationship between recovery fidelity and energy consumption?
    \end{itemize}

    We will answer \textbf{Q1} by presenting our DECA framework in the following, and we address \textbf{Q2} when we evaluate the performance of DECA in Sec.~\ref{sec:perf}. The general idea of DECA is a decomposition between CS aggregation and the recovery of the surveyed field. More specifically, we first recover $\{\mathbf{u}^r\}_{r \in \mathcal{R}}$ from CS coded samples $\{\mathbf{v}^r\}_{r \in \mathcal{R}}$, then $\{F^r\}_{r \in \mathcal{R}}$ is further recovered from $\{\mathbf{u}^r\}_{r \in \mathcal{R}}$. The classic CS technique is used for the first level, but we propose specific diffusion wavelets to act as the sparse basis. This allows an arbitrary network topology for data sampling, as well as virtually any tree partitions to reduce the traffic load. The second level recovery takes the outcome of the first level as noisy sampling data, and manages to recover the field using MC. It absorbs the errors resulting from the previous level, and thus achieves almost the same accuracy as if the sensor data were fully collected.

  \subsection{First Level Recovery} \label{sec:1level}
    The general idea for this level is to apply CS coding to each $\mathbf{u}^r$ through in-network data aggregation. In other words, what the sink collects at the end of round $r$ is $\mathbf{v}^r = \Phi \mathbf{u}^r$. We will explain how this CS coding is applied on top of routing in Sec.~\ref{sec:routing}. Here we are only concerned with recovering $\{\mathbf{u}^r\}_{r \in \mathcal{R}}$ from $\{\mathbf{v}^r\}_{r \in \mathcal{R}}$. According to the discussion in Sec.~\ref{sec:cs}, we could recover individual $\mathbf{u}^r$ by solving an $\ell_1$-minimization problem
    \begin{eqnarray}
      \underset{\mathbf{w}^r \in \mathbb{R}^n}{\mathrm{minimize}} && \|\mathbf{w}^r\|_{\ell_1} = \sum_i |w^r_i| \label{eq:cs+}\\
      \mbox{subject to} && \|\Phi \Psi \mathbf{w}^r - \mathbf{v}^r \|_{\ell_2} \le \epsilon, \nonumber
    \end{eqnarray}
    to obtain the optimal solution $\mathbf{\hat{w}}^r$ and by letting $\mathbf{\hat{u}}^r = \Psi \mathbf{\hat{w}}^r$. The reason we use an error bound as the constraint is due to the fact that real world data are not strictly sparse but just compressible. As this problem is well known and can be solved by various techniques including, for example, $\ell_1$-\textit{magic} \cite{L1magic} and \textit{gradient projection for sparse  reconstruction}(GPSR) \cite{GPSR}, the art of our design lies in constructing $\Psi$.

    \subsubsection{Basis for Spatial Correlation} \label{sec:spatial}
      As $\mathbf{u}^r$ is sampled by a WSN with nodes randomly deployed, the basis $\Psi$ has to adapt to this irregularity, and hence diffusion wavelet basis is an ideal choice. According to Sec.~\ref{sec:diffwave}, diffusion wavelets are generated by diffusion operator $O$, and $O$ in turn comes from the weighted adjacency matrix $\Omega=[\omega_{i,j}]$. As $\omega_{i,j}$ represents the correlation between the data sampled at nodes $i$ and $j$, it should be a function of distance if we want to represent the spatial correlation. Let $d(i,j)$ be the Euclidean distance between nodes $i$ and $j$,  we define
      \begin{eqnarray}
        \omega_{i,j} =
        \begin{cases}
          d^\alpha(i,j) & i \not= j, \\
          \beta & \mathrm{otherwise},
        \end{cases}
        \label{eq:omega}
      \end{eqnarray}
      where $\alpha < 0$ and $\beta$ is a small positive number. As a result, the normalized Laplacian becomes
      \begin{eqnarray}
        \lambda_{i,j} =
        \begin{cases}
          1 - \frac{\beta}{\sum_{p}{d^\alpha(i,p)}}& i=j, \\
          - \frac{d^\alpha(i,j)}{\sqrt{\sum_{p}{d^\alpha(i,p)} \sum_{p}{d^\alpha(p,j)}}} & \mathrm{otherwise}.
        \end{cases}
        \label{eq:lap+}
      \end{eqnarray}
      Here the constant $\beta$ is used to tune the spectrum of the graph $G$, hence the structure of diffusion wavelets.
      \begin{prop} \label{prop1}
        The eigenvalues of $\Lambda$ lie between 0 and 2, and the maximum eigenvalue $\sigma_\mathrm{max}(\Lambda)$ is a decreasing function in $\beta$.
      \end{prop}
      The proof is postponed to Appendix~\ref{apx:prop1}.

\begin{figure}[htb]
        \parbox{\columnwidth}{\parbox{.49\columnwidth}{\center\includegraphics[width=.49\columnwidth]{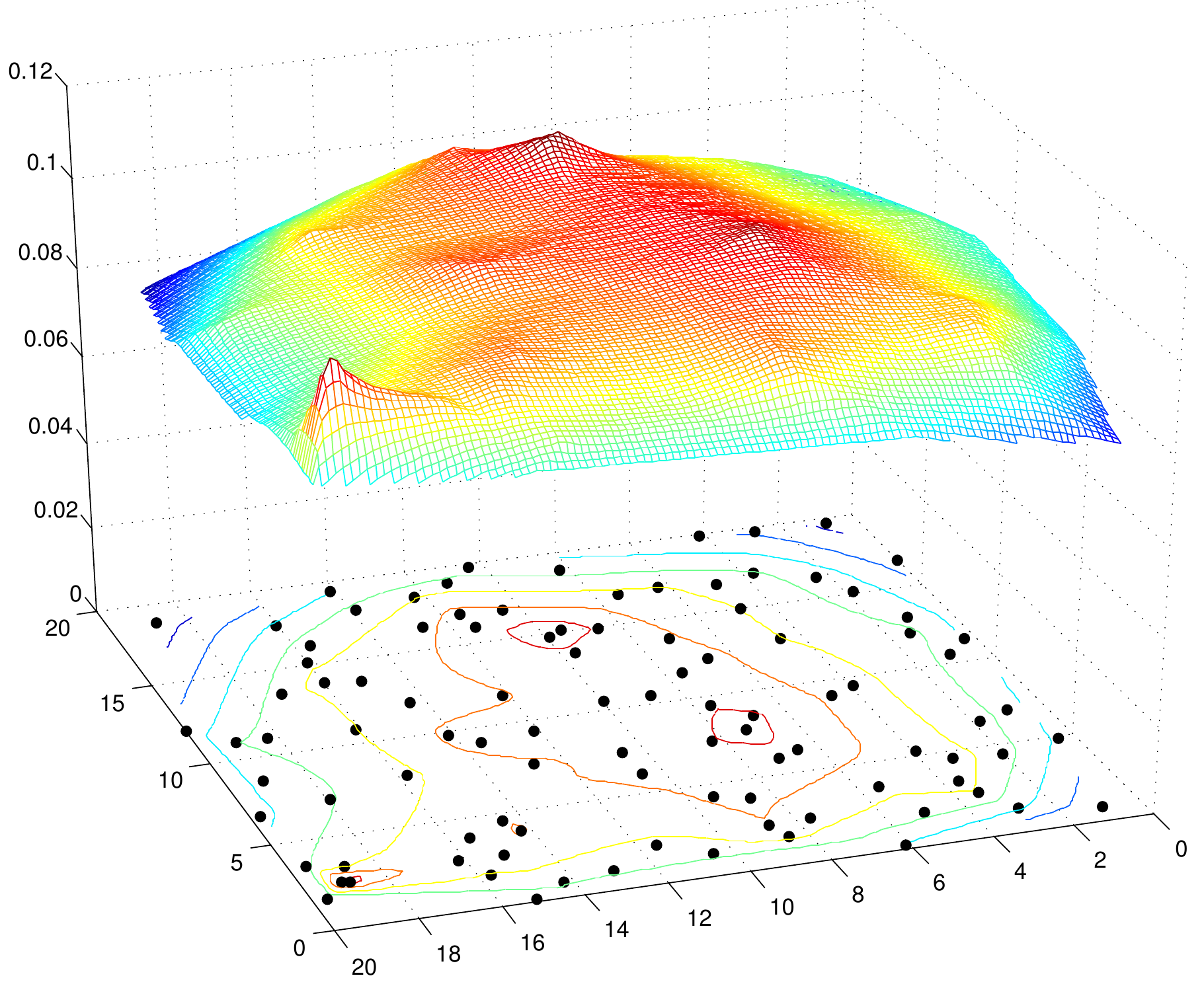}}
                              \parbox{.49\columnwidth}{\center\includegraphics[width=.49\columnwidth]{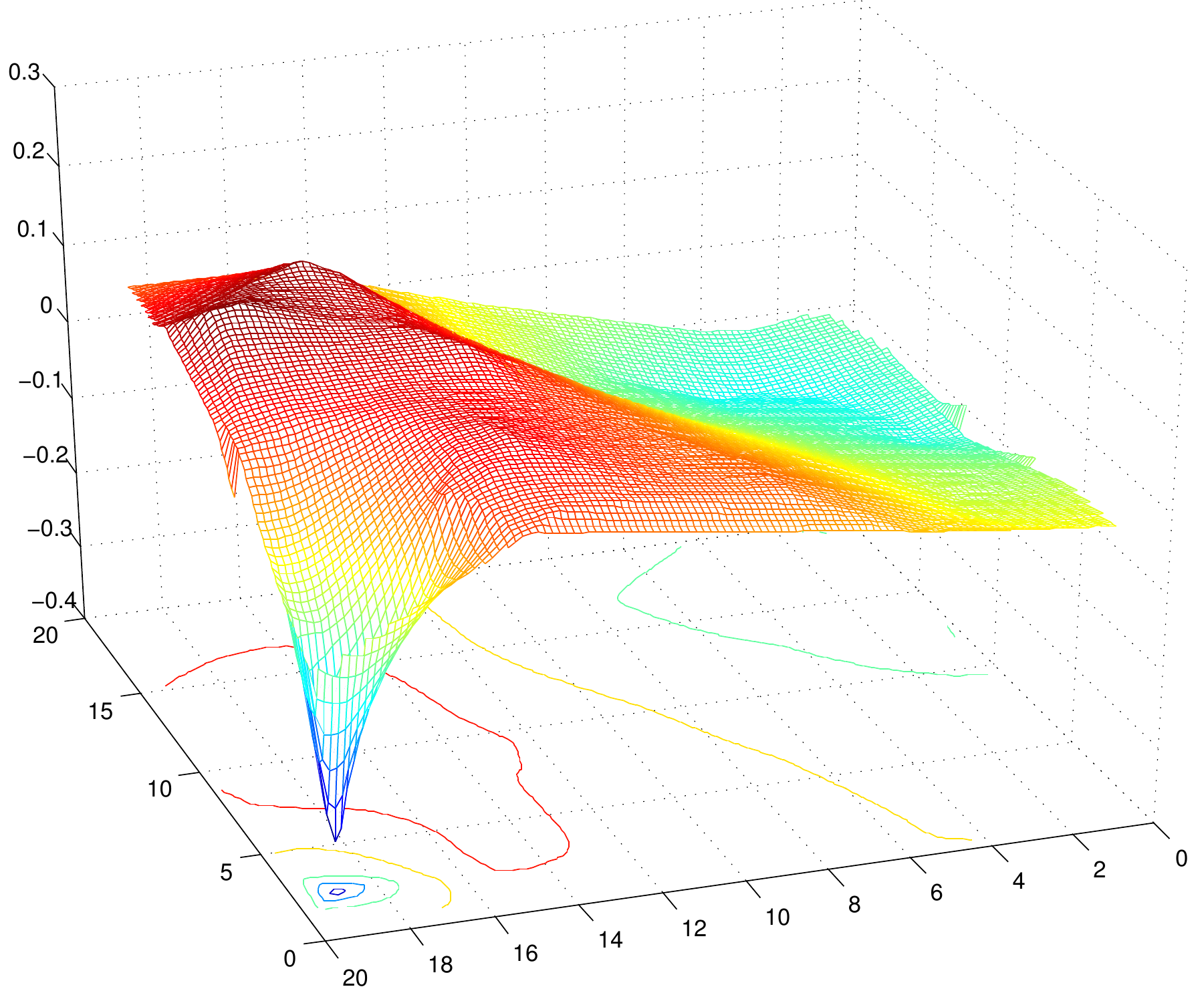}}}
        \parbox{\columnwidth}{\parbox{.49\columnwidth}{\center\scriptsize(a)~Scaling function}
                              \parbox{.49\columnwidth}{\center\scriptsize(b)~Wavelet function I}}
        \parbox{\columnwidth}{\parbox{.49\columnwidth}{\center\includegraphics[width=.49\columnwidth]{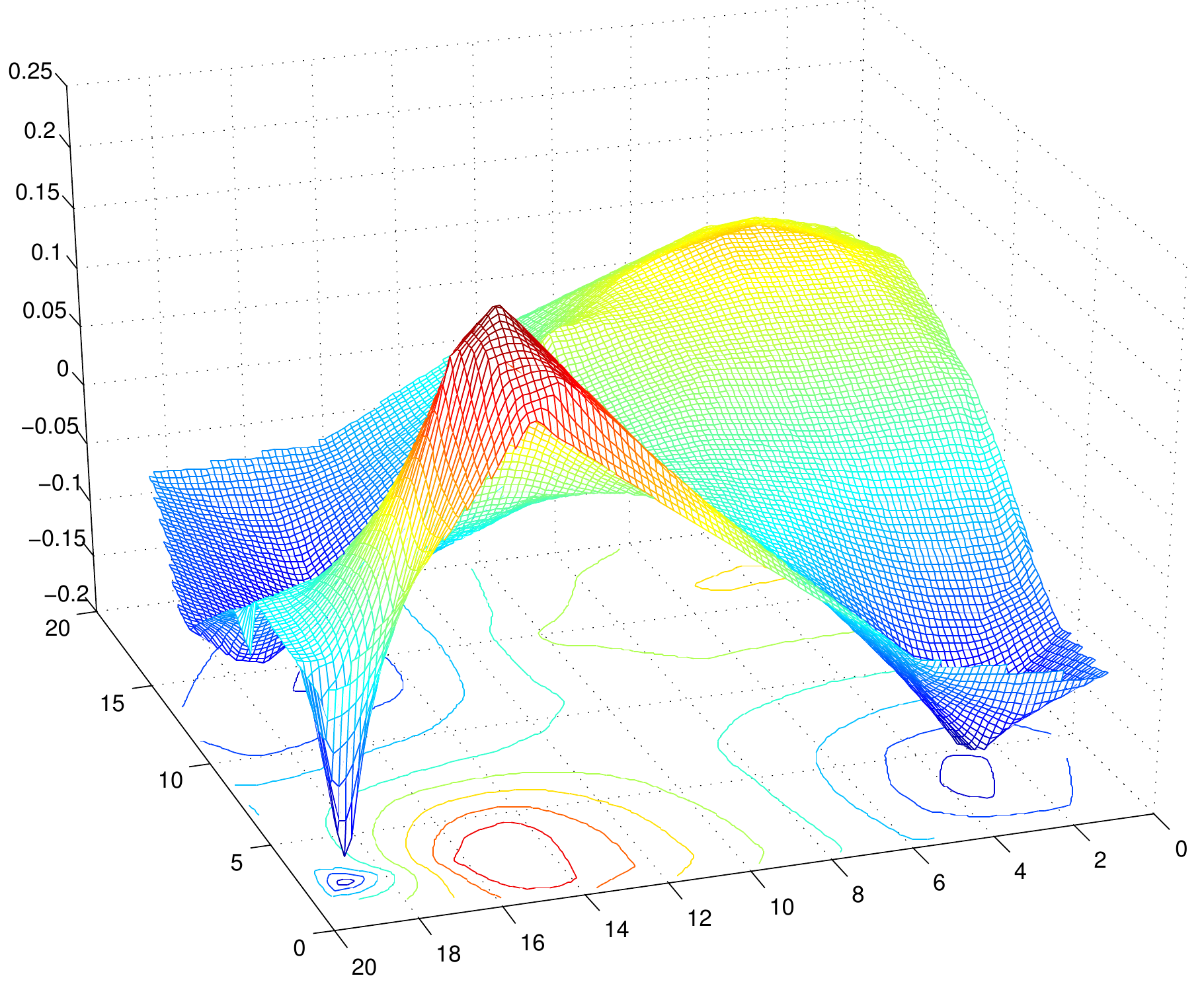}}
                              \parbox{.49\columnwidth}{\center\includegraphics[width=.49\columnwidth]{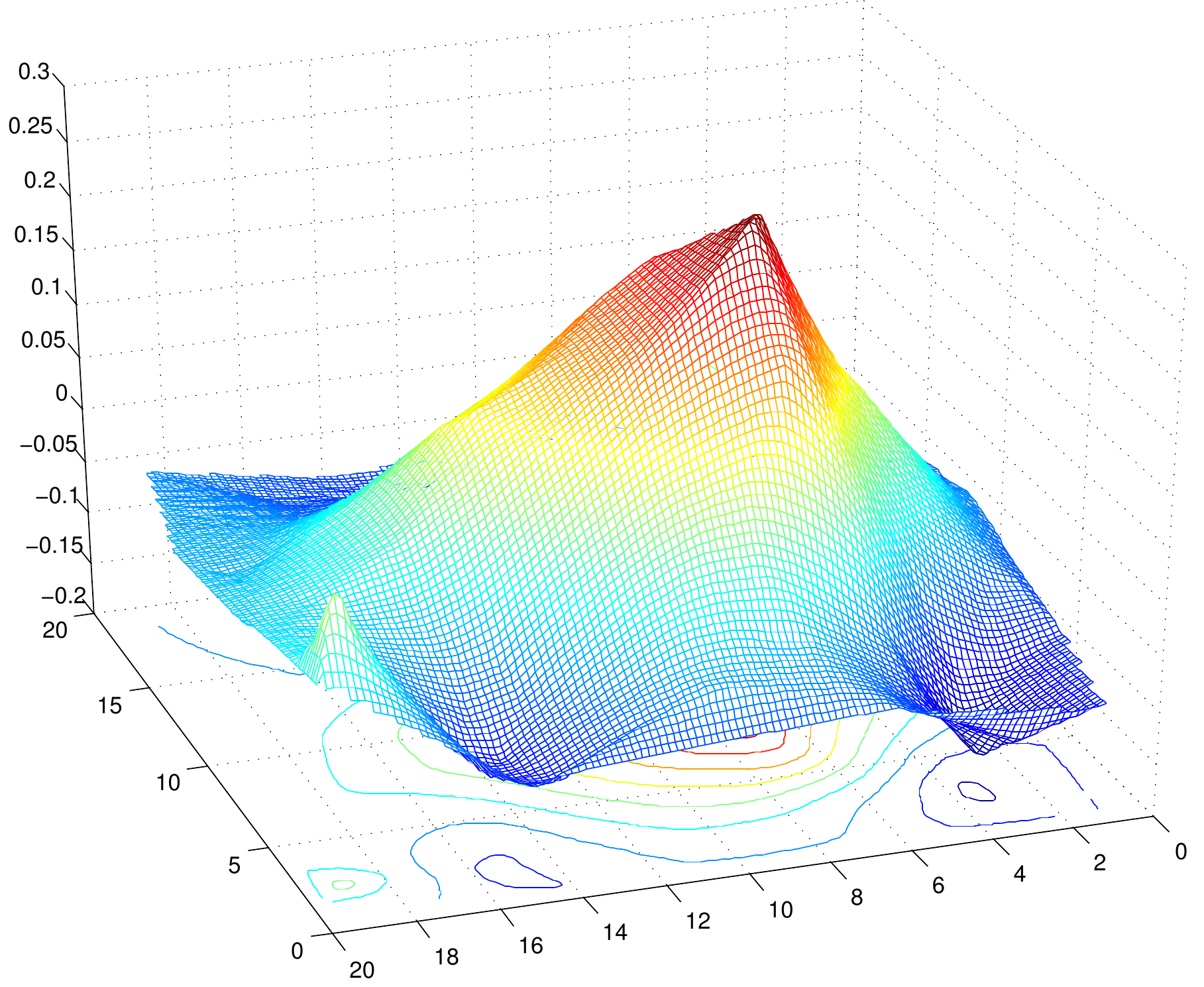}}}
        \parbox{\columnwidth}{\parbox{.49\columnwidth}{\center\scriptsize(c)~Wavelet function II}
                              \parbox{.49\columnwidth}{\center\scriptsize(d)~Wavelet function III}}
        \caption{Diffusion wavelets for a 100-node WSN. Although these wavelets are discrete, we present them as interpolated surfaces to facilitate visual illustration.} \label{fig:dwave}
      \end{figure}

      Based on this proposition, two straightforward choices of the diffusion operator $O$ are ($I$ is the identity matrix):
      \[O = I - \Lambda~~~\mathrm{or}~~~O = \Lambda/2;\]
      both have their eigenvalues ranged from 0 to 1. Therefore, keeping raising a dyadic power to $O$ will make all the eigenvalues diminish eventually. So we partition the range space of $O$ and group the eigenvectors to form the basis, by thresholding on the diminishing eigenvalues of $O^{2^j}$. Based on the above construction procedure, we generate the diffusion wavelets for a WSN with 100 nodes and illustrate some of them in Fig.~\ref{fig:dwave}.

    \subsubsection{Joint Spatial and Temporal Correlation} \label{sec:jstc}
      As the diffusion wavelet basis stems from a graph that represents the data correlation, we can extend the basis defined for spatial correlations to include temporal correlations as well. The idea is that, for a given time period indexed by $\mathcal{R}$, we replicate the graph $G$ by $|\mathcal{R}|$ times and also index them by $\mathcal{R}$. Within each graph $G^r$, the weighted adjacency matrix is still $\Omega$ in (\ref{eq:omega}). Between two graphs $G^{r_1}$ and $G^{r_2}$, the weight between node $i$ in $G^{r_1}$ and node $j$ in $G^{r_2}$ is given by
      \[\omega_{i,j}^{r_1,r_2} = \omega_{i,j} g(|r_1 - r_2|),\]
      where $g(\cdot)$ is an increasing function. This extended adjacency matrix is given as
      \begin{eqnarray}
        \tilde{\Omega} = \!\!
                \left[
                  \begin{array}{cccc}
                    \Omega & \!\!\Omega I_{g(|r_1 - r_2|)} & \cdots & \Omega I_{g(|r_1 - r_{|\mathcal{R}|})}\!\! \\
                    \!\!\Omega I_{g(|r_2 - r_1|)} & \Omega & \ddots & \Omega I_{g(|r_2 - r_{|\mathcal{R}|})}\!\! \\
                    \vdots & \ddots & \ddots & \vdots \\
                    \!\!\Omega I_{g(|r_{|\mathcal{R}|} - r_1|)} & \!\!\Omega I_{g(|r_{|\mathcal{R}|} - r_2|)} & \cdots & \Omega
                 \end{array}
               \right], \nonumber
      \end{eqnarray}
      where $I_{g(|r_1 - r_2|)}$ is a diagonal matrix with $g(|r_1 - r_2|)$ on its diagonal. The temporal correlation represented by $\tilde{\Omega}$ can be fine-tuned by $g(\cdot)$: the larger the first-order derivative of $g(\cdot)$, the faster the temporal correlation attenuates. Based on this extension, we can derive the diffusion operator and hence diffusion wavelets following exactly the same procedure as presented in Sec.~\ref{sec:spatial}.

      Intuitively, the benefit of involving spatial and temporal correlations together is twofold. Firstly, for large-scale WSNs with hundreds or thousands of nodes, the number of measurements $k$ to be collected for each round could be reduced while still achieving the same level of recovery fidelity. Secondly, for small scale WSNs with only tens of nodes, CS often fails to deliver satisfactory recovery accuracy for individual rounds, which stems from the asymptotic nature of the CS theory. However, we could still apply CS aggregation but recover sensory data from a certain period (several consecutive rounds) as a whole. We will confirm these intuitions in Sec.~\ref{sec:intel}.

  \subsection{Second Level Recovery} \label{sec:2level}
    Based on our assumption in Sec.~\ref{sec:netmodel}, the field on top of the whole sensing area (which is only sparsely covered by a WSN) can be deemed as an ``image'' with each cell being a ``pixel'' of the image. Therefore, given the recovery results $\mathbf{\hat{u}}^r$ from the first level, one could use the CS technique again: solve another $\ell_1$ minimization problem (with $\mathbf{\hat{u}}^r$ as the input) to recover the field. As this time the data items are well aligned into a 2D grid, $\Psi$ can be any basis for image compression, including DCT or wavelets.

    However, it is well known that SVD is optimal in decomposing a matrix into separable (additive) components, as it leads to the minimum number of coefficients. Mathematically, for a matrix $X$, we have $X = \sum_{i=1}^{\mathrm{rank}(X)}\sigma_i A_i$, where $\sigma_i$ is the $i$-th singular value and $A_i$ is a rank-1 matrix given by the outer product of the $i$-th left and right singular vectors. Therefore, a better choice is to use $\{A_i\}$ as a counterpart of $\Psi$ for ``matrix CS''. This indeed corresponds to the idea of matrix completion discussed in Sec.~\ref{sec:mc}. Unfortunately, we cannot directly solve the nuclear-norm minimization problem (\ref{eq:mc}), because in general $\mathbf{\hat{u}}^r \not= \overrightarrow{\mathcal{P}_\Pi(F^r)}$, where $\Pi$ refers to the location of WSN nodes in the matrix, $\mathcal{P}_\Pi$ is the operator defined in Sec.~\ref{sec:mc}, and $\overrightarrow{\cdot}$ transforms a matrix to a vector indexed by node IDs. As the original sensory data $\mathbf{u}^r$ is often not sparse but only compressible, $\mathbf{\hat{u}}^r$ is only an approximation of $\mathbf{u}^r$ in both $\ell_1$ and $\ell_2$ norms \cite{Candes-SPMag08}. Therefore, we have $\mathbf{\hat{u}}^r = \overrightarrow{\mathcal{P}_\Pi(F^r)} + \boldsymbol{\xi}$, with $\boldsymbol{\xi}$ being the error term resulting from the previous level, and the current recovery relies on solving another optimization problem
    \begin{eqnarray}
      \mathrm{minimize} && \|X\|_\ast \label{eq:mc+} \\
      \mbox{subject to} && \|\mathbf{\hat{u}}^r - \overrightarrow{\mathcal{P}_\Pi(X)}\|_{\ell_2} \le \delta. \nonumber
    \end{eqnarray}
    Since the error bound $\delta$ comes from the assumption that $\|\boldsymbol{\xi}\|_{\ell_2} \le \delta$, it depends on the parameters of the first level recovery (e.g., $\epsilon$ and $m$). According to the theory of stable matrix completion \cite{CandesP-ProcIEEE10}, the optimal solution of this problem approximates $F^r$ in $\ell_2$ norm.

    As a summary of the joint effect of the dual-level recovery, we have the following result.
    \begin{prop} \label{prop2}
      If $n \ge \mathcal{O}(\max(a,b) \zeta \log^\tau\!(\max(a,b)))$ (second level) and $k \ge \mathcal{O}\left(m\log(n/m)\right)$ (first level), the optimal solution of (\ref{eq:mc+}), using the optimal solution of $(\ref{eq:cs+})$ as input, approximates $F^r$ in $\ell_2$ norm:
      \[ \|\hat{X} - F^r\|_2 \le \left(4\sqrt{\frac{(2ab+n)\min(a,b)}{n}} + 2\right)\delta.\]
    \end{prop}
    This is proven by combining the results stated in \cite{Candes-SPMag08} and \cite{CandesP-ProcIEEE10}, as shown in Appendix~\ref{apx:prop2}. We will show in Sec.~\ref{sec:perf} that, in practice, $k/(ab)$ is below 5\%. This means that, to recover a field represented by $a\times b$ samples, only less than 5\% measurements need to be collected from the monitoring WSN.

  \subsection{Efficient Data Routing} \label{sec:routing}
    The routing design for DECA is on top of the hybrid CS aggregation scheme introduced in Sec.~\ref{sec:csrt}. As we demonstrated in \cite{XiangLV-SECON11}, better energy efficiency can be achieved with decreasing of $k$. However, applying CS aggregation on the whole network with a small $k$ might not suggest an acceptable recovery. Thanks to the large redundancy in sensory data, intuitively, we can expect its spatially-localized subset is still sparse in a proper domain. Then for large-scale WSNs, we seek to further cut down the energy cost by partitioning the network into several subnetworks and carrying out CS coding (with a small $k$) independently within each part. At the decoding end, we take the joint reconstruction~\cite{LeePSKO-GSN09} to recover the data. Formally, the entire aggregation structure is a set $\mathcal{T}$ of disjoint data aggregation trees, all rooted at distinct one-hop neighbors of the sink and each tree $\mathcal{T}_i \in \mathcal{T}$ has an $n_i < n, k_i < k$ with $\sum_{\mathcal{T}_i \in \mathcal{T}} n_i = n, \sum_{\mathcal{T}_i \in \mathcal{T}} k_i \approx k$. By tuning $k_i$, we hope to strike a balance between the energy efficiency and recovery performance. Note that each $\mathcal{T}_i$ is constructed to be nearly optimal using the greedy algorithm presented in \cite{XiangLV-SECON11}. However, the partition causes the sensing matrix $\Phi$ to have a block-diagonal shape:
    \begin{eqnarray}
      \Phi = \left[
                \begin{array}{cccc}
                  \Phi_1 & \mathbf{0} & \cdots & \mathbf{0} \\
                  \mathbf{0} & \Phi_2 & \ddots & \mathbf{0} \\
                  \vdots & \ddots & \ddots & \vdots \\
                  \mathbf{0} & \mathbf{0} & \cdots & \Phi_{|\mathcal{T}|}
               \end{array}
             \right], \nonumber
    \end{eqnarray}
    where $\Phi_i$ is a $k_i \times n_i$ matrix with random entries as specified in Sec.~\ref{sec:cs}. Now the question is whether this $\Phi$ satisfies RIP (which requires $\Phi$ to be full; see Sec.~\ref{sec:cs}). Fortunately, we have the following result:
    \begin{prop} \label{prop3}
       For a given signal $\mathbf{u}=\Psi\mathbf{w}$ with $\|\mathbf{w}\|_{\ell_0}=m$, and a partition scheme as stated above, $\mathbf{u}$ can be recovered exactly with high probability from the random samples $\mathbf{v}=\Phi\mathbf{u}$ by solving (\ref{eq:cs+}) if the number of samples satisfies \[k=\mathcal{O}(m|\mathcal{T}|\log^2n).\]
    \end{prop}
The proof is based on the results provided in \cite{LeeO-ASC10} (\textit{Proposition 3.3} and \textit{Theorem 3.4}). Readers are referred to Appendix~\ref{apx:prop3} for a sketch.

    %

\section{Performance Evaluations} \label{sec:perf}
  In this section, we evaluate the performance of DECA with respect to recovery accuracy and energy efficiency, based on a large number of experiments using both synthetic data and real data sets. Moreover, we also address \textbf{Q2} stated in Sec.~\ref{sec:netmodel}; we show that DECA allows a WSN user to fine-tune the tradeoff between recovery accuracy and energy efficiency.

  \subsection{Experiment Settings}
    Existing online data sets are often collected by small-scale WSNs (e.g., EPFL SensorScope \cite{Sensorscope} and Intel Labs Berkeley WSN data \cite{Intel}). To also mimic widespread fields monitored by large-scale WSNs, we come up with three different ways to generate the data sets for our experiments.
    \begin{enumerate}
      \item \textbf{Peak}: A synthetic data set generated by \textsf{peaks} function in Matlab.
      \item \textbf{Intel}: Real data sets obtained by a WSN deployed at Intel Labs Berkeley \cite{Intel}.
      \item \textbf{Temp}: Temperature distribution in USA retrieved from \url{http://www.weather.gov}.
    \end{enumerate}

    For the first and the last data sets, we take a subset spread on a square area with $100\times100$ cells. The (field) value within each cell is set to be constant. In order to monitor such a field, we deploy a WSN on it by randomly putting nodes in cells with a coverage rate $\rho$, so the network size is $n=\rho\cdot10^4$. We fix $\rho$ for the first data set, but we will vary it for the third data set. The in-network CS aggregation is performed in two ways: it either routes data through a single tree with sample size $k$, or through four disjoint trees of equal size, with $k_i = k/4$ or $k/3$ for each tree.

    For the first level recovery of the multi-tree CS aggregated data, we either apply the diffusion wavelet basis for the whole WSN to directly recover the sensory vector $\mathbf{u}$, or we conduct CS recoveries for individual trees using their respective diffusion wavelet bases. We call these two mechanisms \textit{joint recovery} (JR) and \textit{independent recovery} (IR), respectively. We use $\ell_1$-magic \cite{L1magic} and FPC \cite{MaGC-MP09} to solve the minimization problems for the first and second level recoveries. The performance of recovery accuracy is measured by the \textit{recovery error}, defined as the \textit{normalized mean square error} in the following:
    \begin{eqnarray}
      \varepsilon = \frac{\|\mathbf{u} - \mathbf{\hat{u}}\|_{\ell_2}}{\|\mathbf{u}\|_{\ell_2}}& \mathrm{or}&\varepsilon = \frac{\|X - \hat{X}\|_2
          }{\|X\|_2}.
    \end{eqnarray}
    They are defined for vector recovery (first level) and matrix recovery (second level), respectively. For energy efficiency evaluation, we set the single-hop transmission cost $c(i,j)$ to be proportional to the cubic of the distance between the communicating pair $i$ and $j$.

    During our experiments, we have tested upon numbers of diffusion operators, by varying $\alpha$ and $\beta$ in the Laplacian (\ref{eq:lap+}) and by setting $O = I-\Lambda$ or $O = \Lambda/2$. According to our observations, $O=I-\Lambda$ performs much better than $O=\Lambda/2$ as the sparse basis, while the parameter tuple $\alpha\in[-1,-1/3]$ and $\beta\in[0,2]$ suggests the sparsest representation for given sensory data (e.g., Fig.~\ref{fig:peakscoef}). Therefore, we fix $\alpha=-1$ and $\beta=1$ and use $O=I-\Lambda$ in later experiments.

    \vspace{1ex}\noindent\emph{Remark: We will not compare DECA with other mechanisms, as DECA is the only one that can handle field recovery based on incomplete data samples.}

  \subsection{Synthetic Field: Peak} \label{sec:peak}
\begin{figure}[ht]
      \begin{center}
        \parbox{.48\columnwidth}{\center \includegraphics[width=.44\columnwidth]{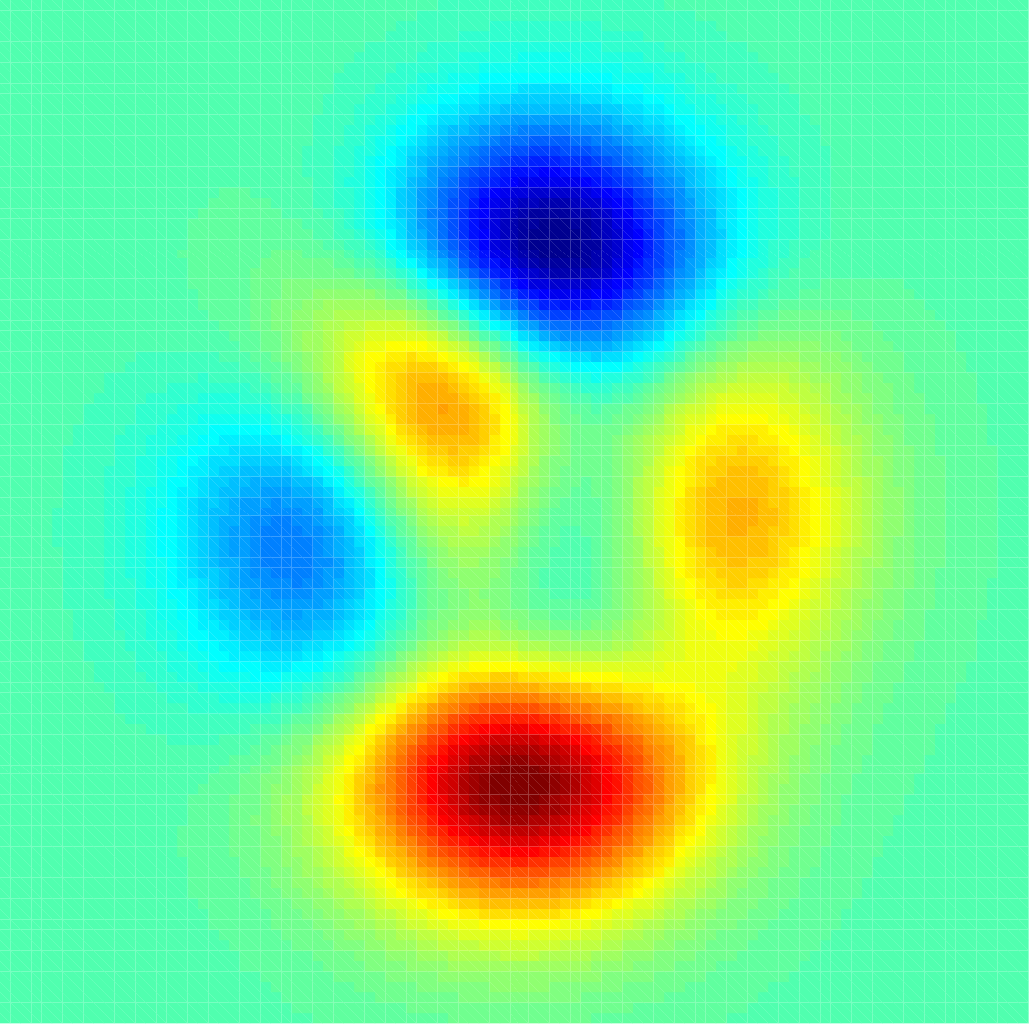}}
        \parbox{.48\columnwidth}{\center \includegraphics[width=.45\columnwidth]{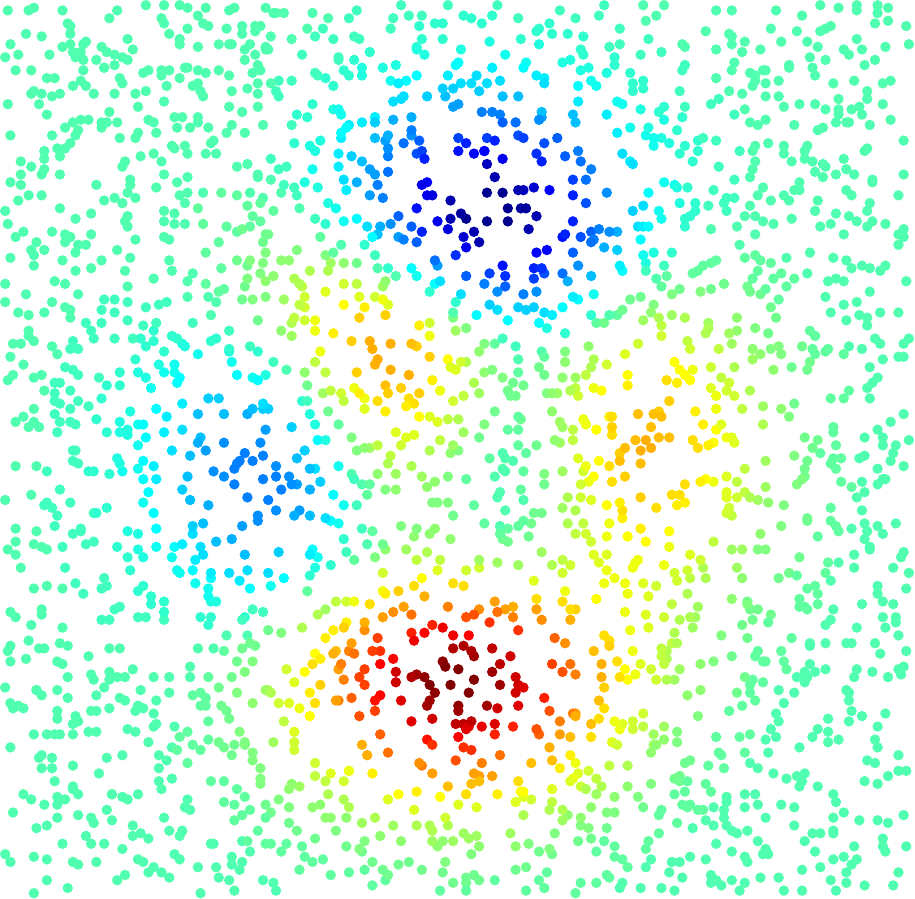}}
        \parbox{.48\columnwidth}{\center\scriptsize~(a) Original field $F$}
        \parbox{.48\columnwidth}{\center\scriptsize~(b) Sensory data $\mathbf{u}$}
        \parbox{.48\columnwidth}{\center \includegraphics[width=.44\columnwidth]{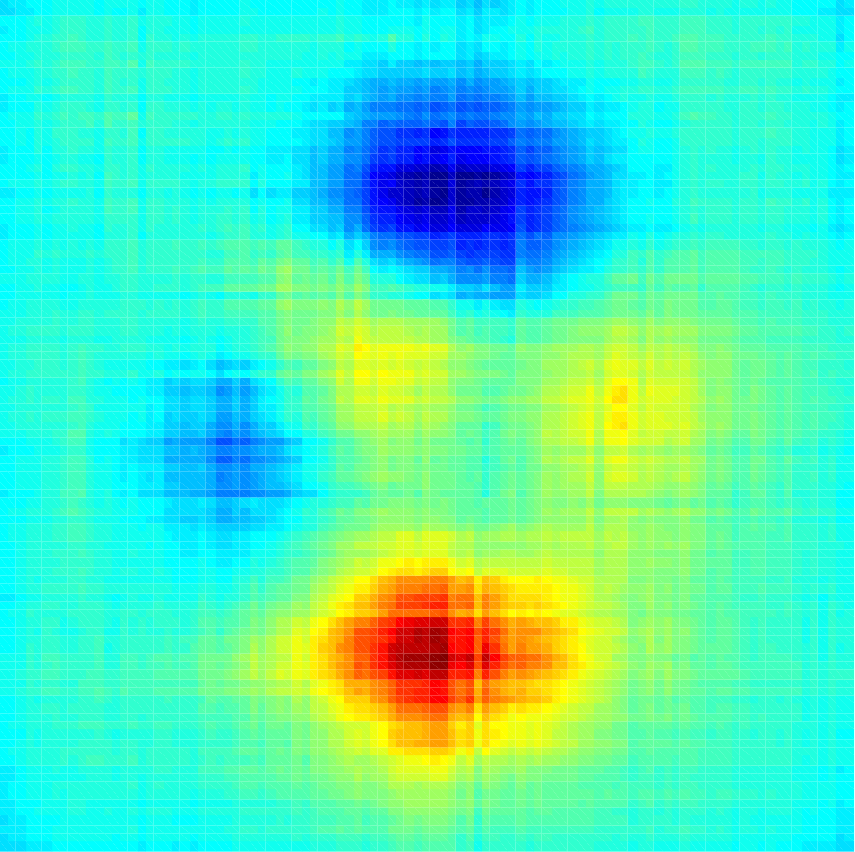}} 
        \parbox{.48\columnwidth}{\center \includegraphics[width=.45\columnwidth]{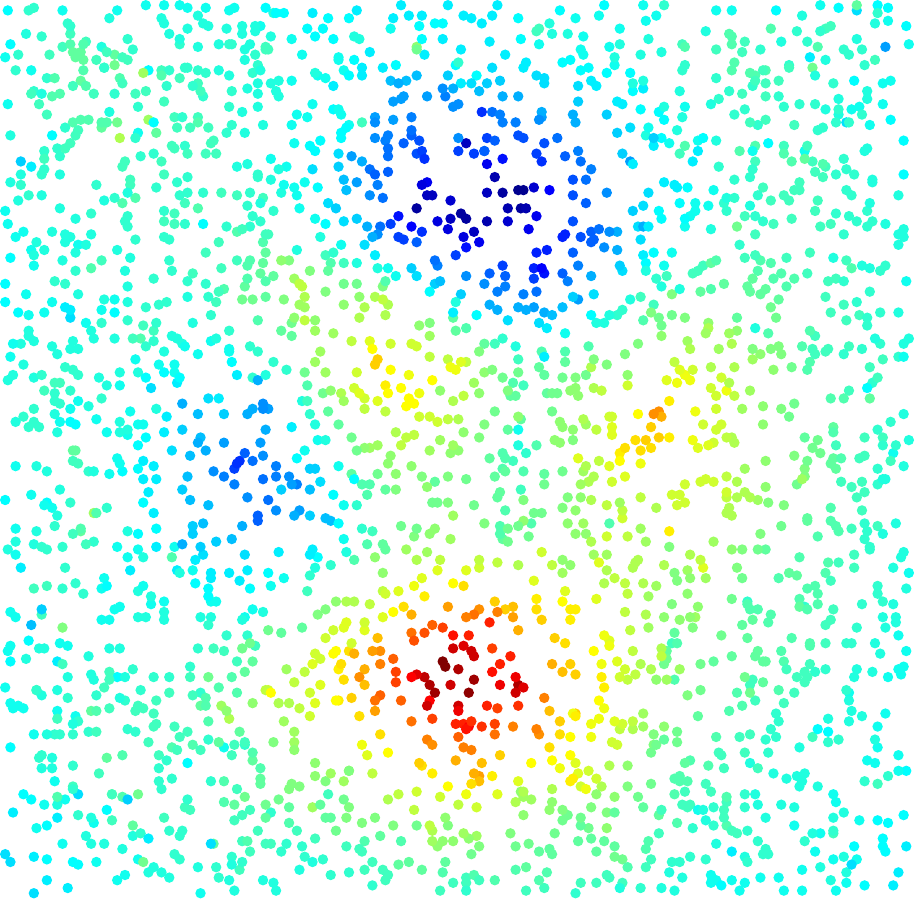}} 
        \parbox{.48\columnwidth}{\center\scriptsize~(d) 2nd level recovery $\hat{X}$}
        \parbox{.48\columnwidth}{\center\scriptsize~(c) 1st level recovery $\mathbf{\hat{u}}$}\\
        \caption{Illustration of the DECA process (in clockwise order). We have $n = 2600$ and $k_i = 120$ for each subtree.}
        \label{fig:peaksdata}
      \end{center}
    \end{figure}

    This field is generated by the \textsf{peaks} function in Matlab, whose 2D image is shown in Fig.~\ref{fig:peaksdata}(a). A WSN is randomly deployed on the field, with the sink fixed at the center to collect data from the whole network. We illustrate the dual-level recovery process of DECA in Fig.~\ref{fig:peaksdata}.
    For later experiments, we run through the DECA process 10 times (with different random CS coding) on each of the 10 random deployments to conduct dual-level recovery, and we report the mean values of these 100 processes.

\begin{figure}[ht]
      \begin{center}
        \includegraphics[width=.9\columnwidth]{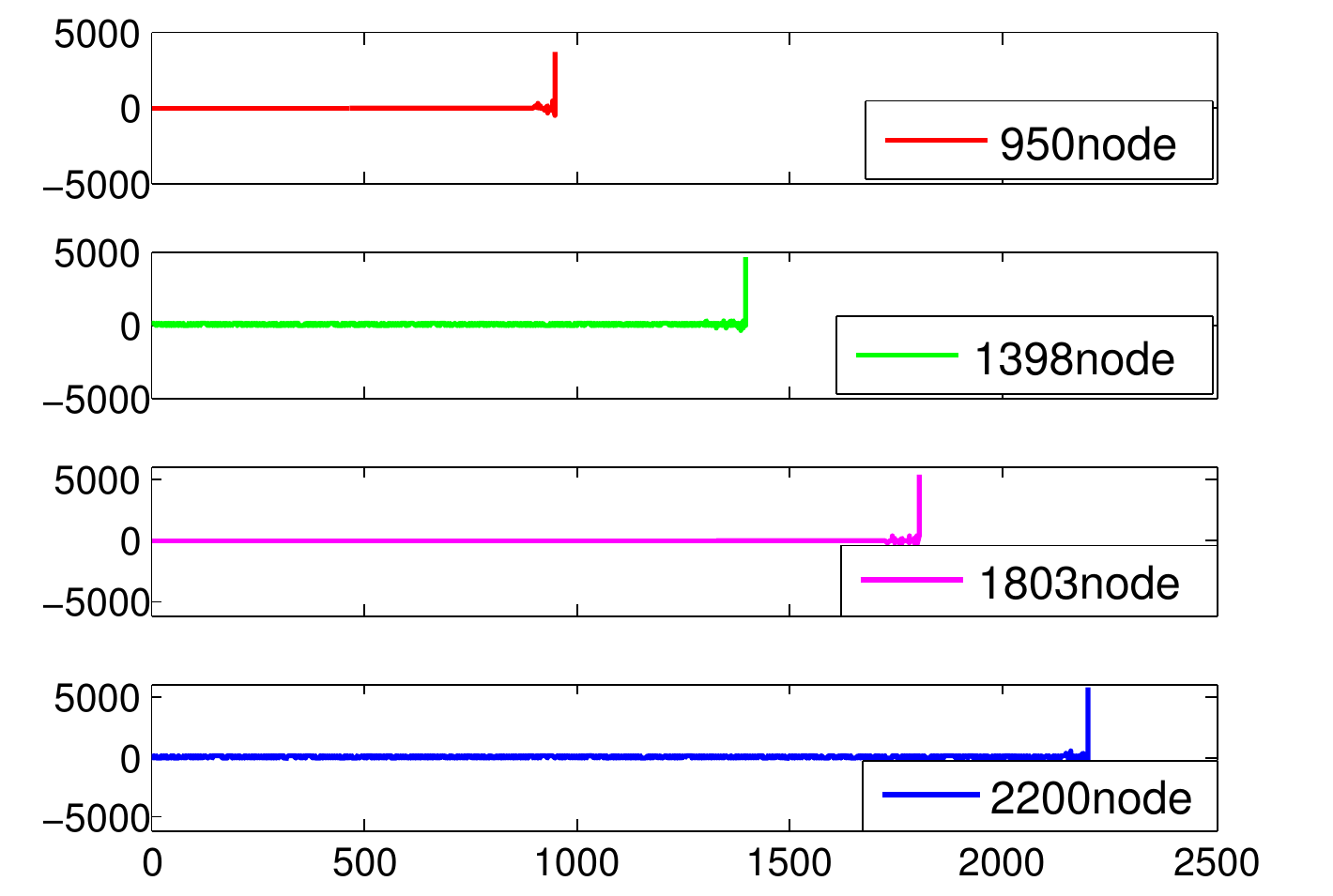}
        \caption{Diffusion wavelet coefficients for WSNs with different sizes.}
        \label{fig:peakscoef}
      \end{center}
    \end{figure}

    In order to show that CS aggregation works well for tree-partitioned WSNs, we need to demonstrate that the sensory data $\mathbf{u}$ only has low frequency components when projected onto the diffusion wavelet basis (see Sec.~\ref{sec:routing}). In Fig.~\ref{fig:peakscoef}, we plot the diffusion wavelet coefficients for sensory data obtained by WSNs of different sizes. As these coefficients are sorted in descending order by their frequencies,
    it is evident that sensory data contain mostly low frequency components of the diffusion wavelet basis.
    \begin{figure*}[t]
      \begin{center}
        \includegraphics[width=.33\textwidth]{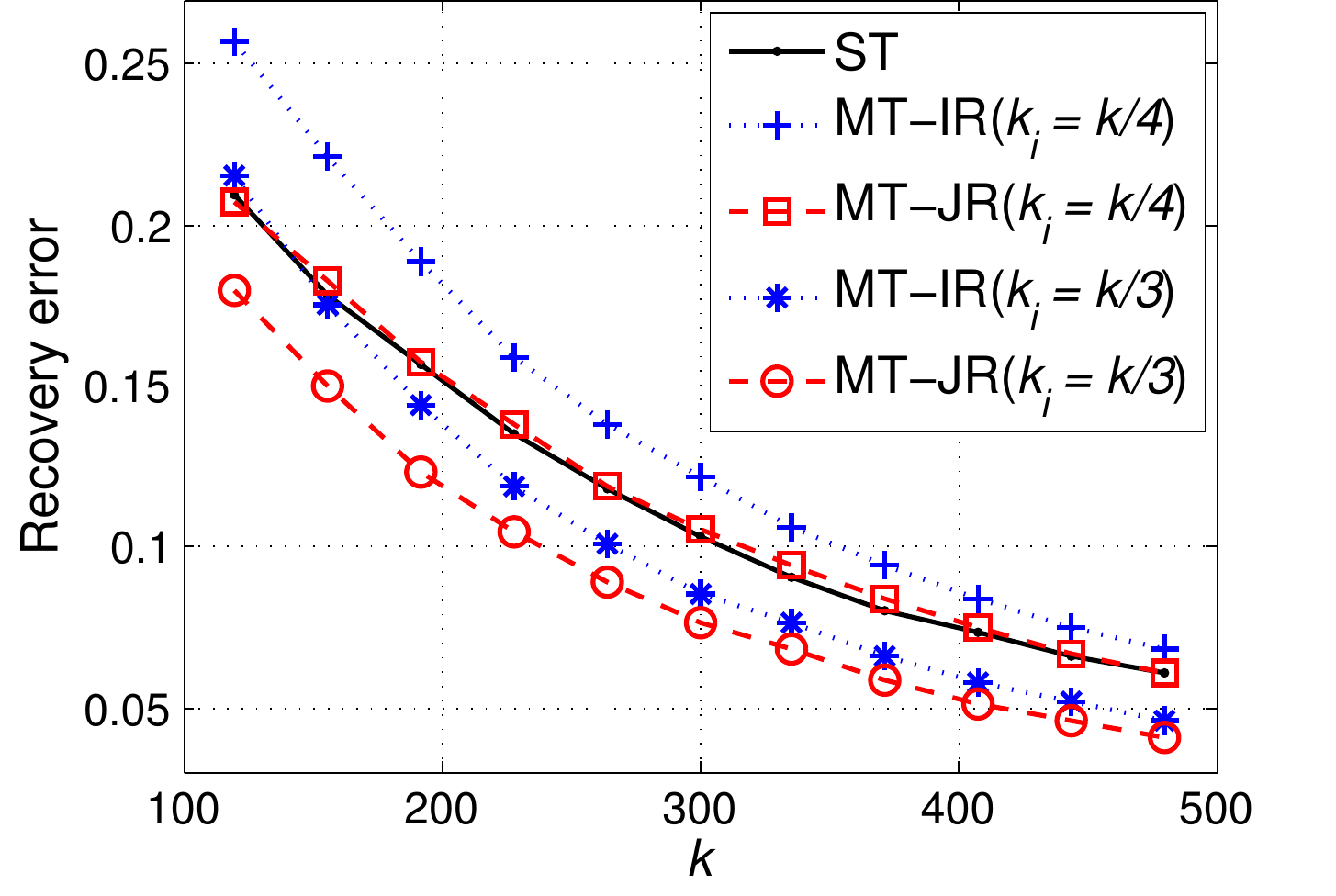}
	    \includegraphics[width=.33\textwidth]{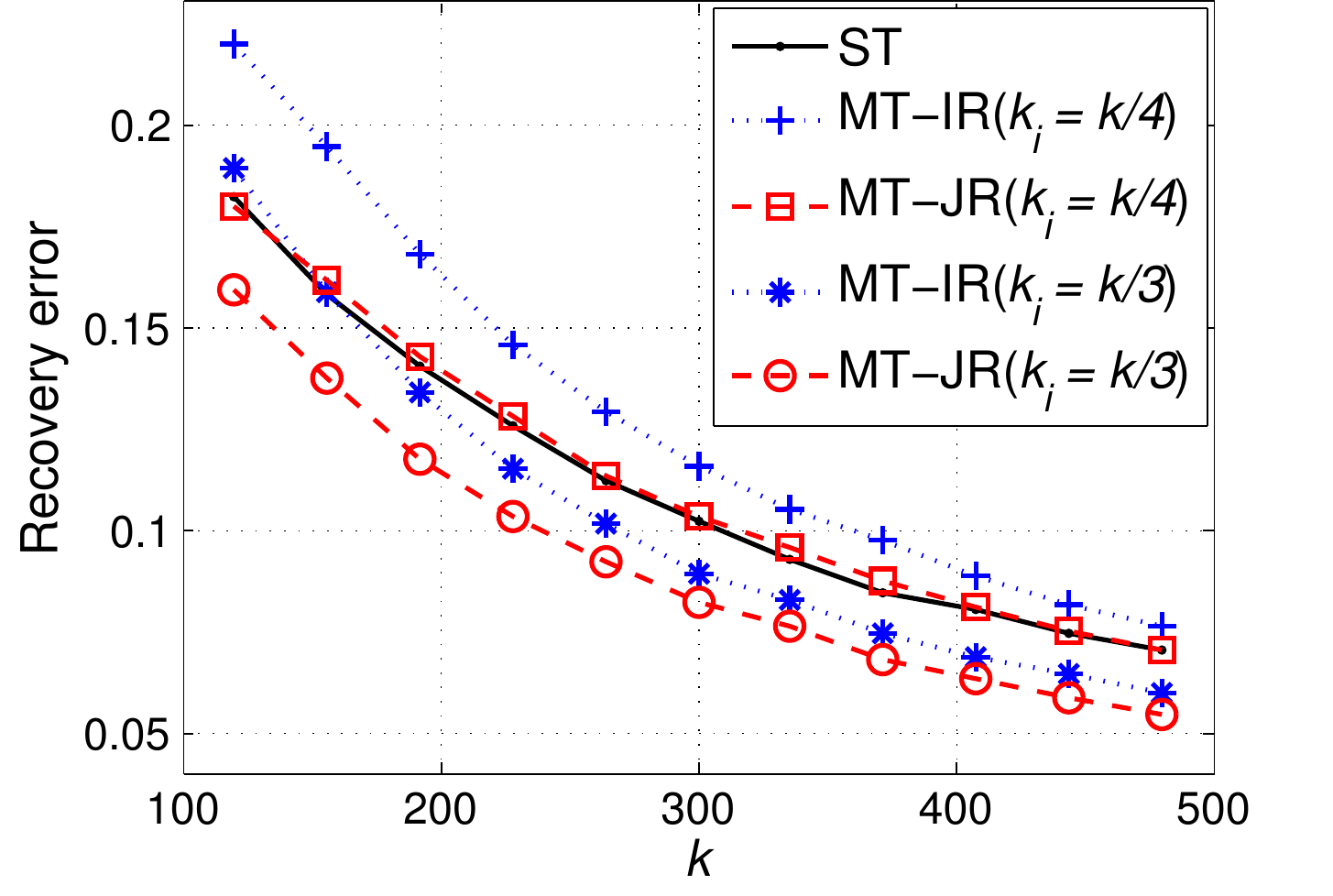}
        \includegraphics[width=.32\textwidth]{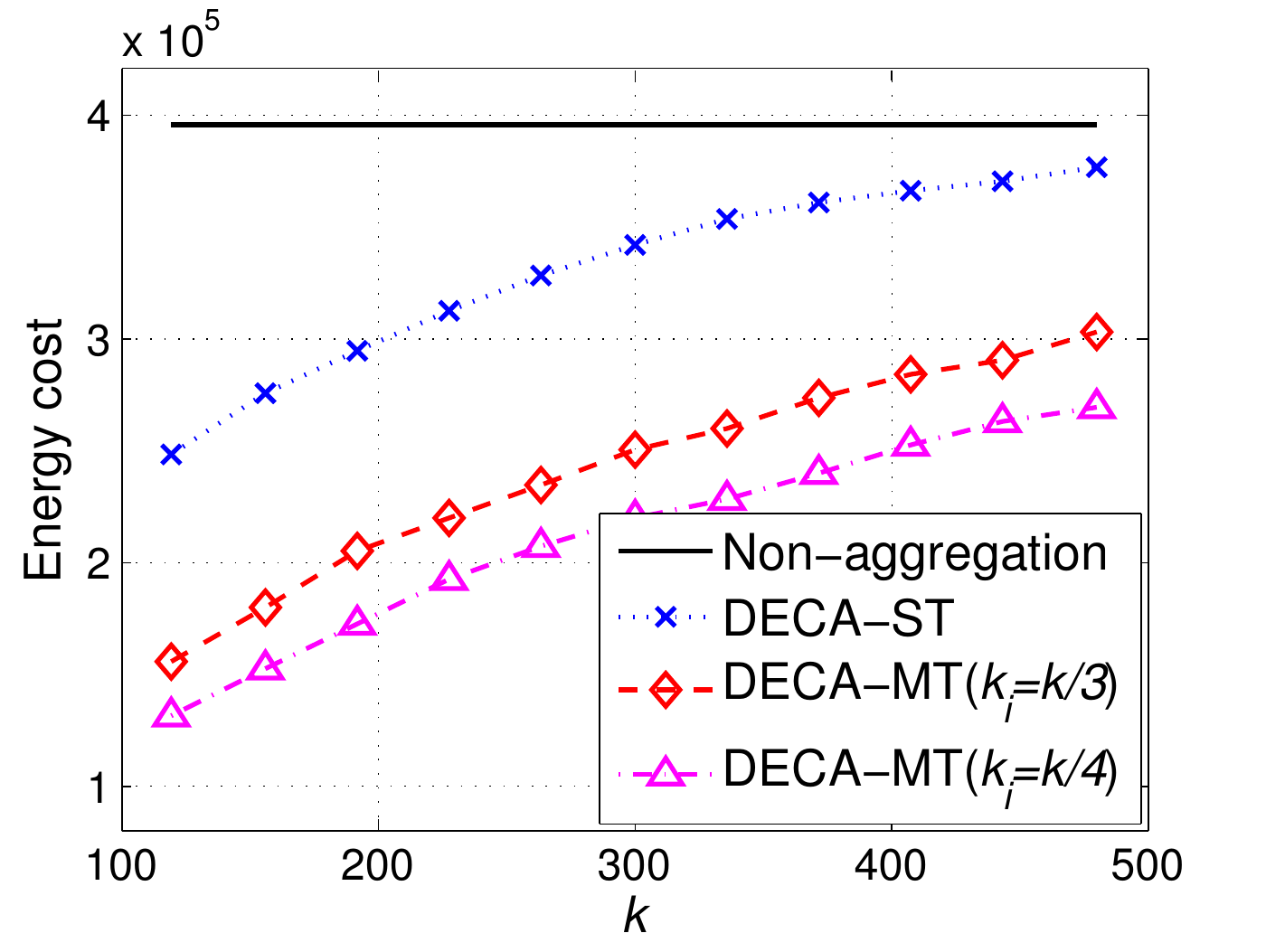}
        \parbox{.33\textwidth}{\center\scriptsize~(a) 1st level recovery error}
        \parbox{.33\textwidth}{\center\scriptsize~(b) 2nd level recovery error}
        \parbox{.32\textwidth}{\center\scriptsize~(c) Aggregation cost}
        \caption{Comparisons based on synthetic data. Here ST/MT corresponds to CDA on single tree or multiple trees. Whereas $k$ refers to the CS measurements in ST case, $k_i$ denotes the measurements used for each subtree. IR and JR are short for independent recovery and joint recovery, respectively.}
        \label{fig:peakscomp}
      \end{center}
    \end{figure*}

    Now we fix the coverage ratio $\rho=0.18$, so the WSN has 1800 nodes. We evaluate the tradeoff between recovery accuracy and aggregation cost, by tuning $k$ (the number of measurements) for CS coding. The results are given in Fig.~\ref{fig:peakscomp}: while (a) and (b) show respectively the first and the second level recovery errors for different data aggregation schemes, (c) plots the energy costs incurred by these schemes.
    If CS coding is performed on a single tree, compared with non-aggregation, around $k=7\%n$ CS measurements lead to lower than 20\% final recovery error and 40\% saving in energy cost. Of course, adding more measurements will continuously reduce the recovery errors at the cost of increasing the aggregation cost.

    To further reduce the energy consumption, we carry out a four-equal-tree partition and conduct CS coding independently within each subtree. If we take $k_i=k/4$ measurements for each subtree, compared with the single tree case, independent recovery leads to worse performance whereas joint recovery gives almost the same outcome. We may attribute this to the loss of correlation between adjacent partitions under independent recovery. In fact, this suggests that, using diffusion wavelet basis, a block diagonal sensing matrix is comparable with a full sensing matrix. Meanwhile, the energy consumption is almost halved.

    To improve the recovery performance, we set $k_i=k/3$, and it performs better than the single tree case in terms of the recovery error, while still reducing at least 40\% of the energy consumption. In summary, DECA, especially with its tree partitioning CS aggregation, achieves very high energy efficiency while preserving the fidelity of data recovery.

    \begin{figure*}[t]
      \begin{minipage}{\textwidth}
      \begin{center}
        \includegraphics[width=.329\textwidth]{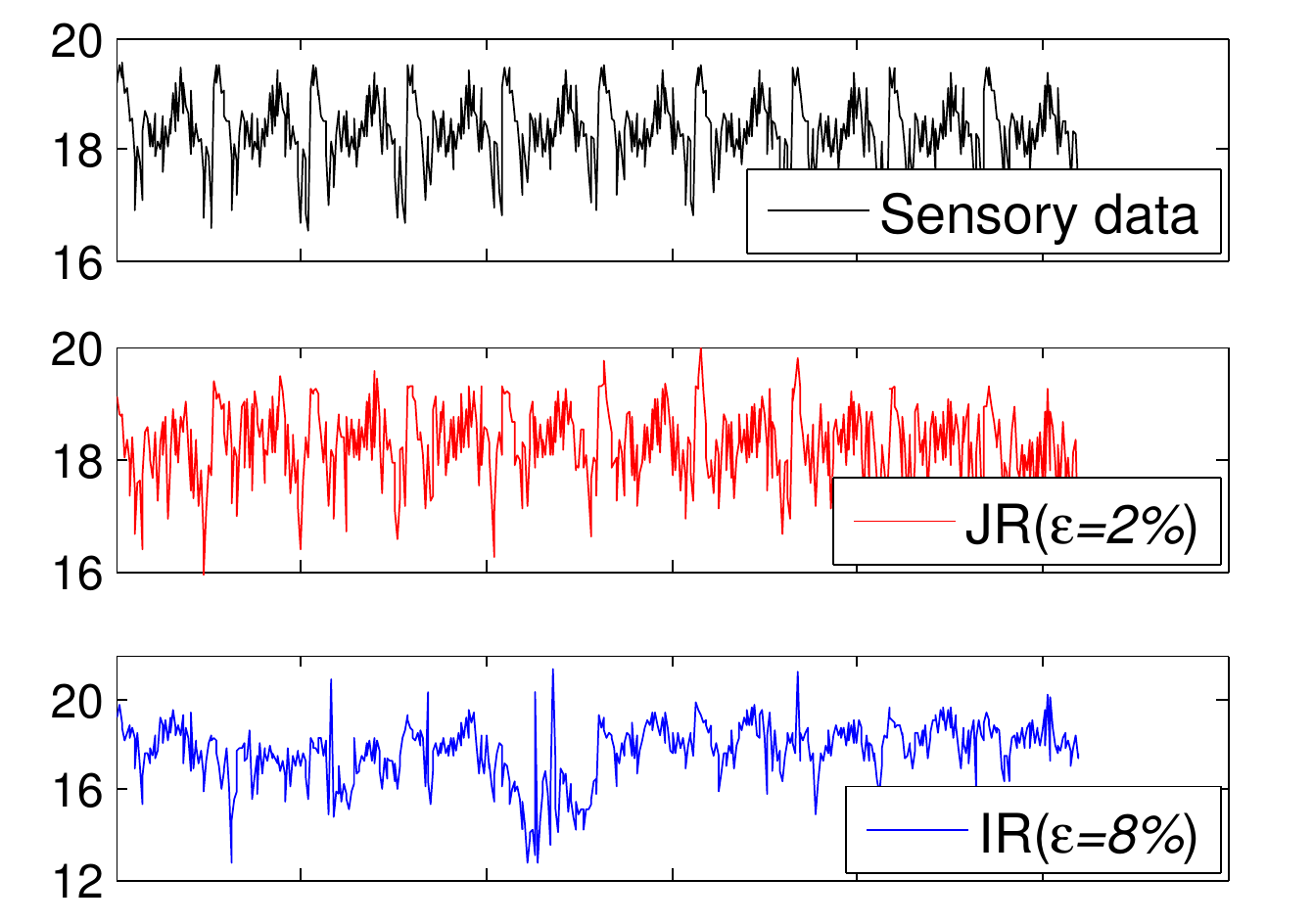}
        \includegraphics[width=.327\textwidth]{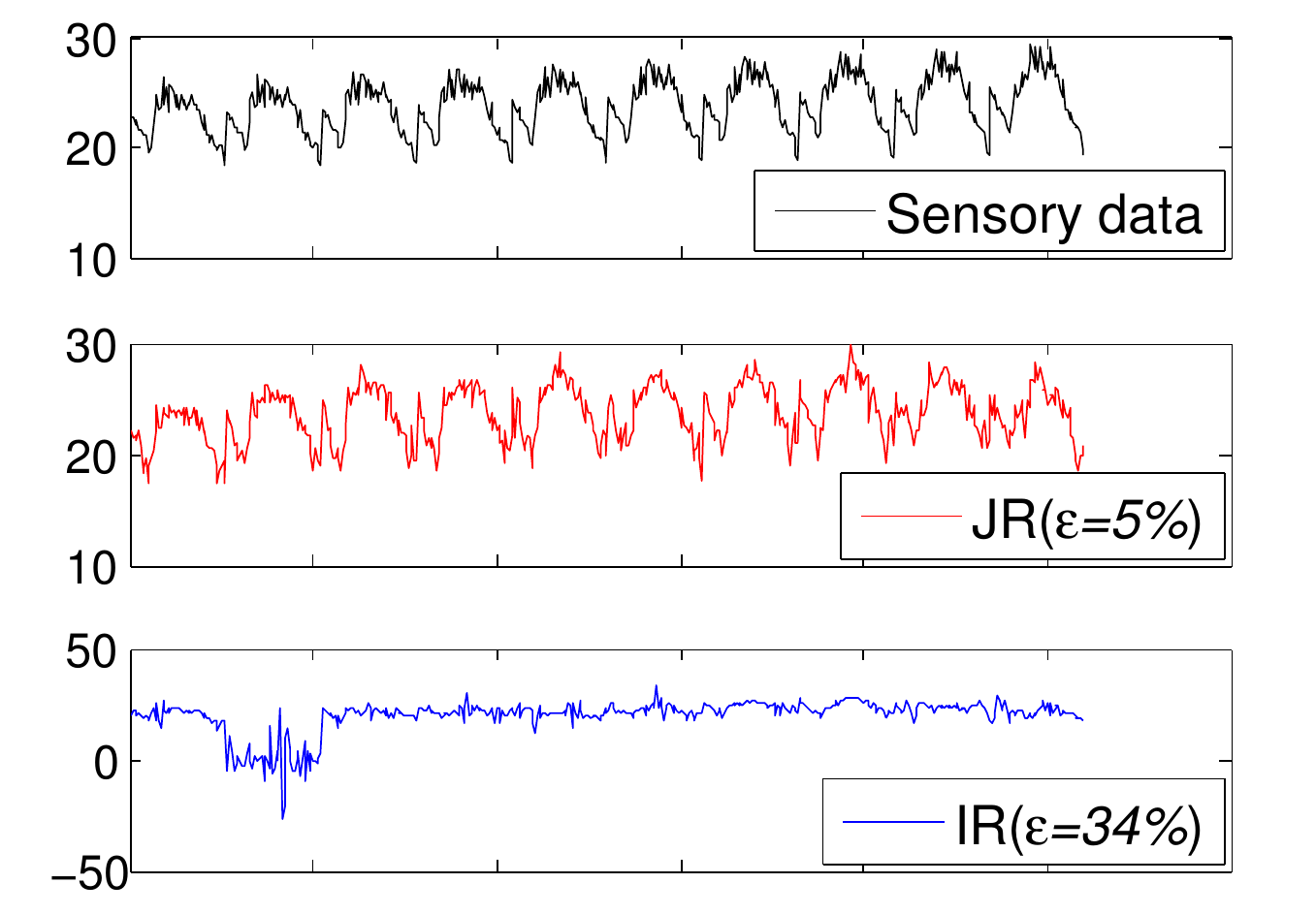}
        \includegraphics[width=.329\textwidth]{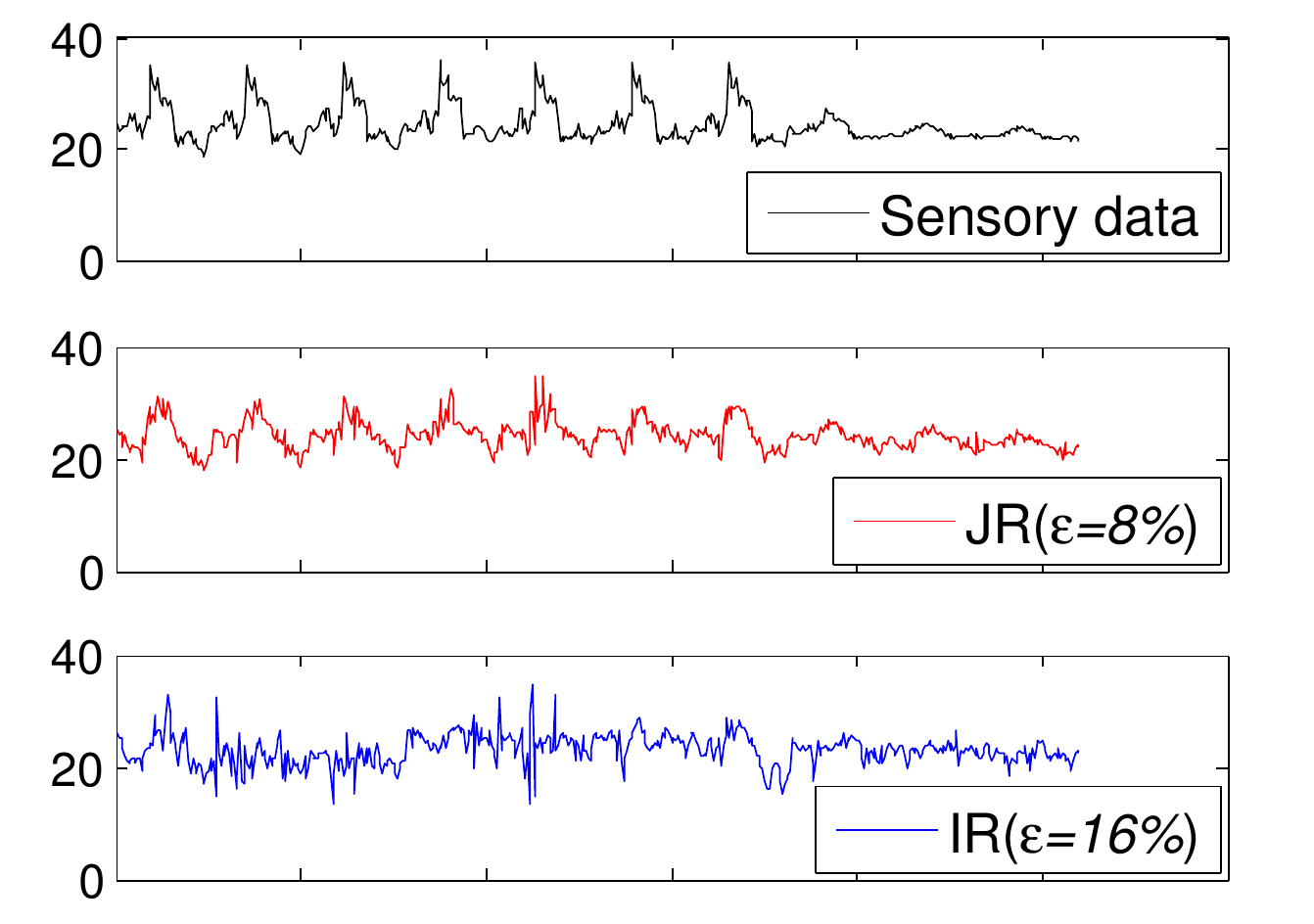}
        \parbox{.329\textwidth}{\center\scriptsize~(a) Every 10min readings}
        \parbox{.327\textwidth}{\center\scriptsize~(b) Every 10min readings}
        \parbox{.329\textwidth}{\center\scriptsize~(c) Every 30min readings}
        \caption{Joint spatial and temporal recovery in a WSN with $n=54$ and $k=10$.}
        \label{fig:temprec}
      \end{center}
      \end{minipage}
      \begin{minipage}{\textwidth}
      \begin{center}
        \includegraphics[width=.33\textwidth]{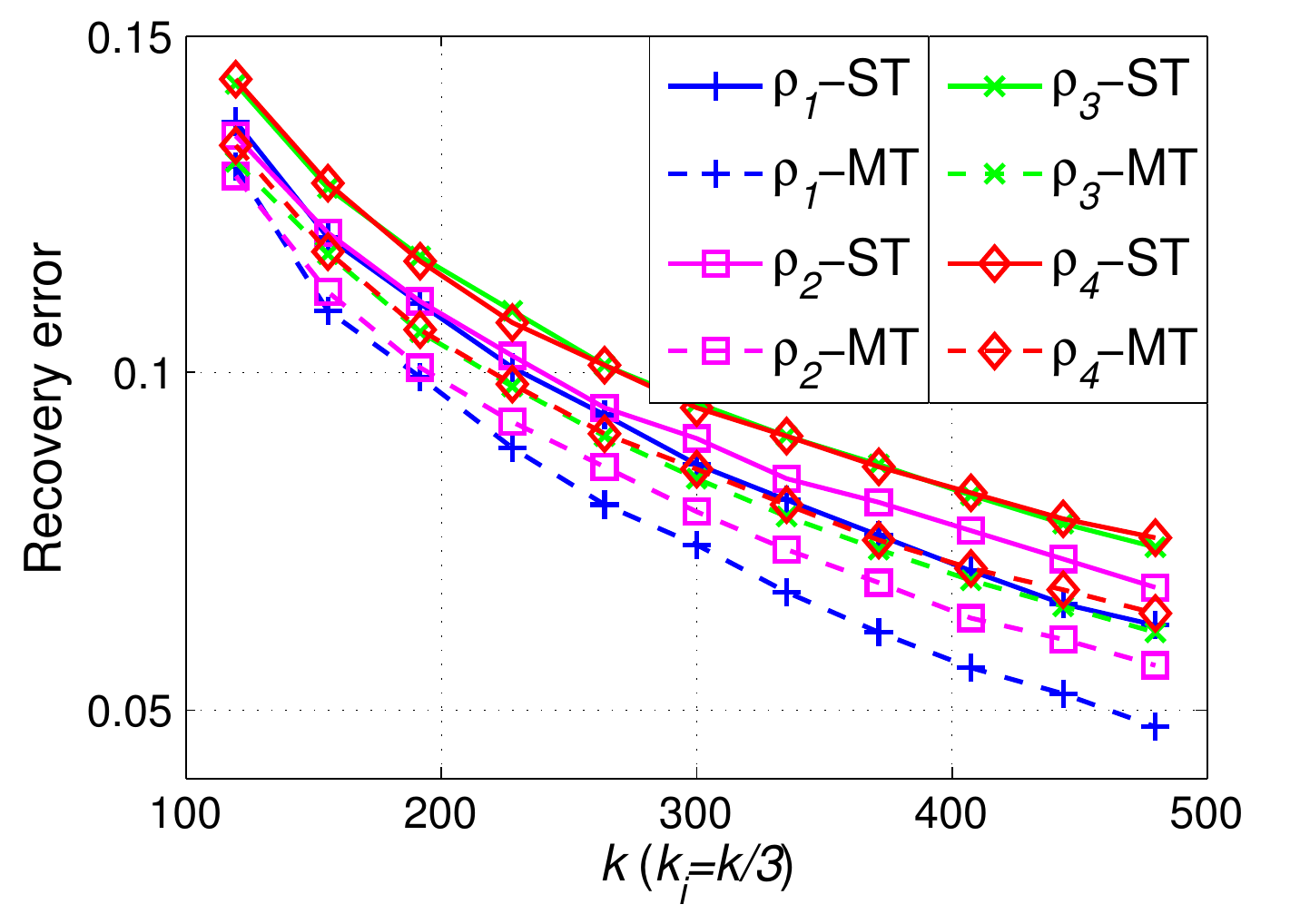}
	    \includegraphics[width=.33\textwidth]{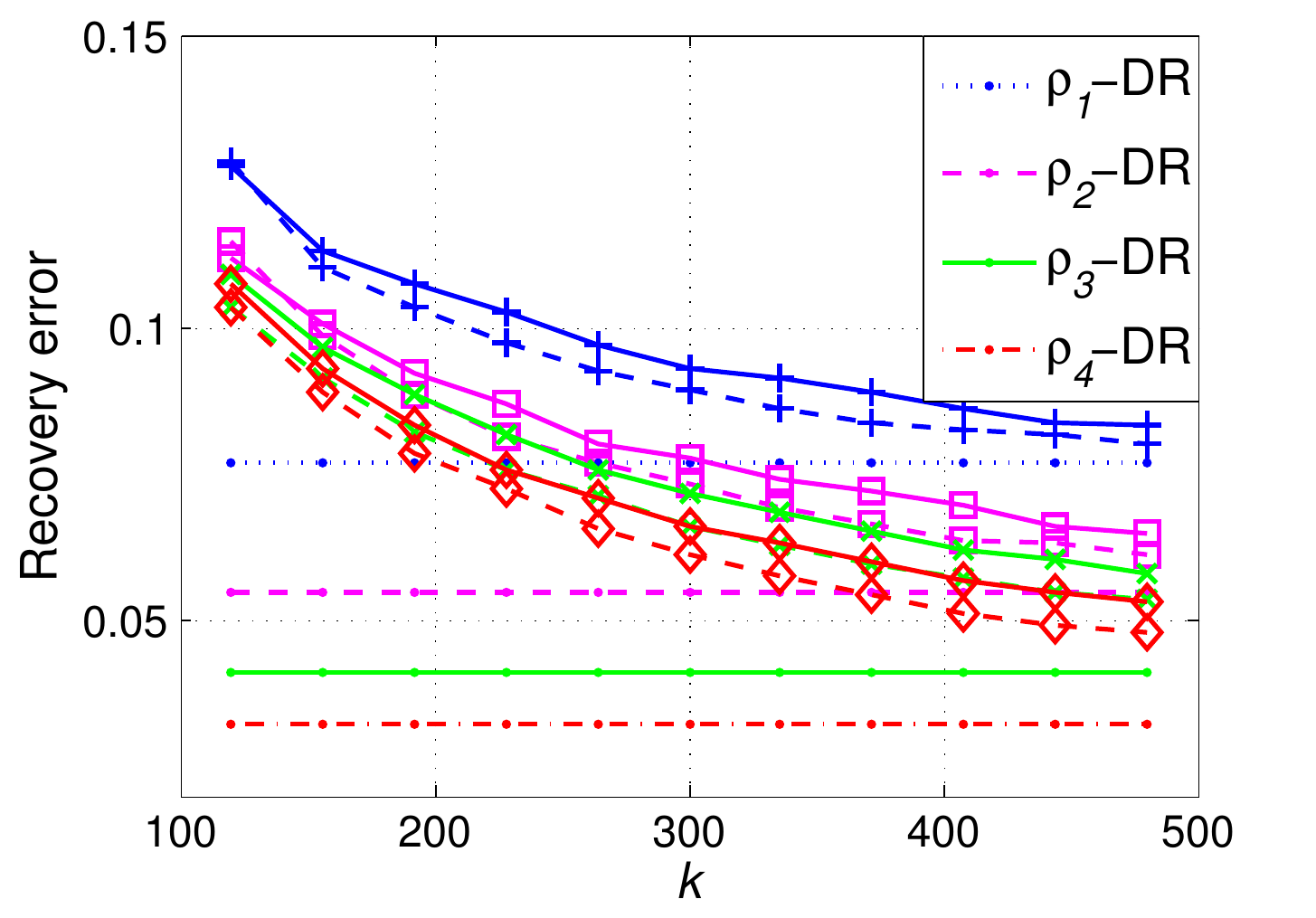}
        \includegraphics[width=.327\textwidth]{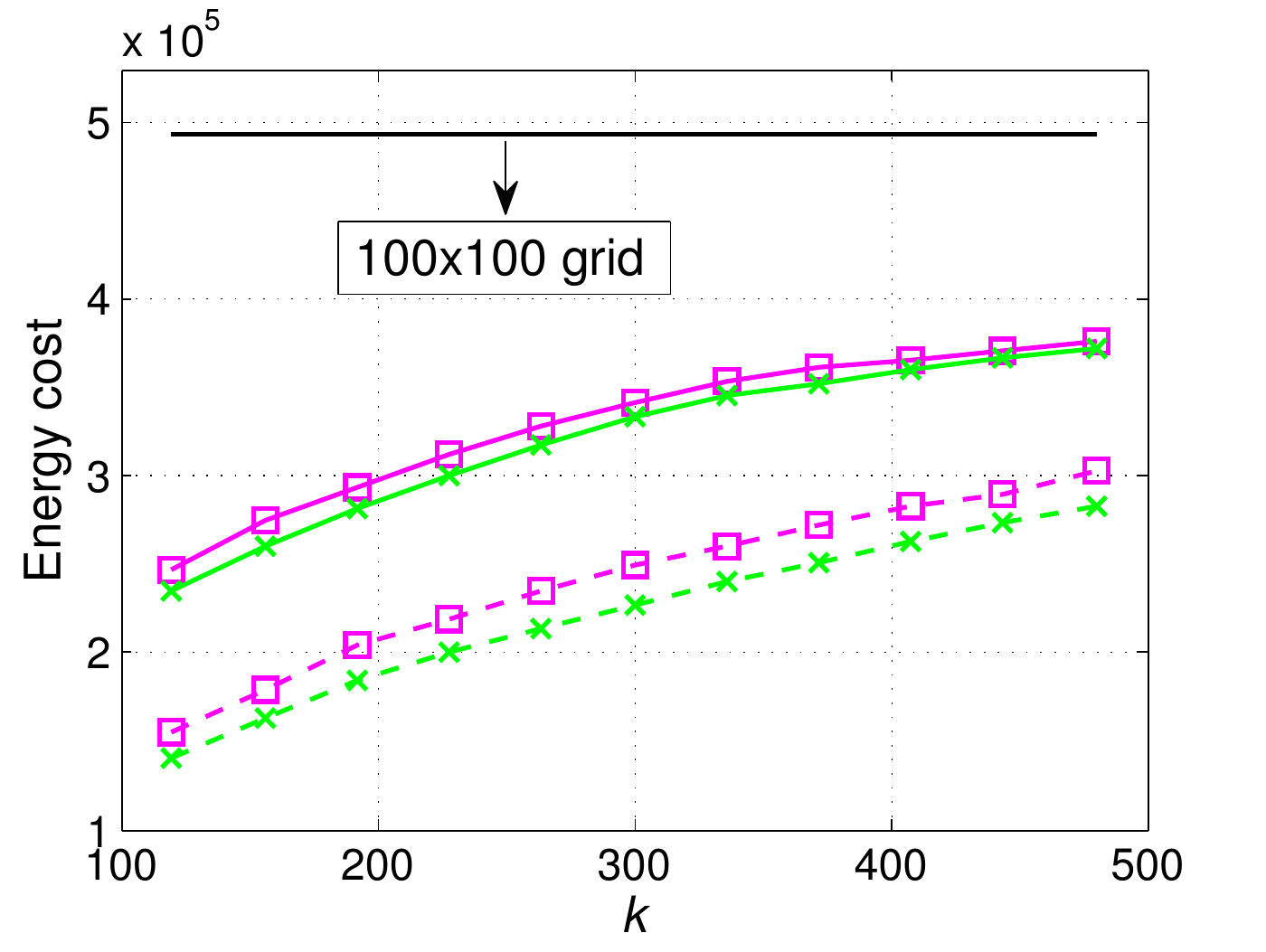}
        \parbox{.33\textwidth}{\center\scriptsize~(a) 1st level recovery error}
        \parbox{.33\textwidth}{\center\scriptsize~(b) 2nd level recovery error}
        \parbox{.327\textwidth}{\center\scriptsize~(c) Aggregation cost}
        \caption{Comparisons based on USA temperature field, where $\rho_1=0.14$, $\rho_2=0.18$, $\rho_3=0.22$, and $\rho_4=0.26$. In all the MT cases, $k_i=k/3$ for each subtree. In (b), DR plottings indicate the direct recoveries from the sensory data, which serve as baselines for the 2nd level performance. As for (c), the baseline is the aggregation cost of a $100\times100$ grid network that fully covers the ``image".}
        \label{fig:weathercomp}
      \end{center}
      \end{minipage}
    \end{figure*}

  \subsection{Real Data from An Actual WSN} \label{sec:intel}
    Now we proceed to analyze DECA on the real sensory data collected by Intel Labs Berkeley~\cite{Intel}. This WSN consisted of 54 sensors. With such a small scale,  it might not be worth applying CS aggregation, as the required number of CS measurements could be comparable to the network size $n$. However, what we want to show here is that, as DECA allows jointly recovering several consecutive snapshots $\{\mathbf{u}^r\}_{r\in\mathcal{R}}$ by leveraging on the temporal correlation, CS aggregation can still be useful for small-scale WSNs.

    To jointly consider spatial and temporal correlations in data, we make use of the diffusion operator proposed in Sec.~\ref{sec:jstc} to generate diffusion wavelet basis. We take 10 consecutive (in time) snapshots, with the time interval between two sets of readings equal to 10 or 30 minutes. As the further two snapshots are away from each other in time, the less likely they are correlated. Therefore, when constructing the diffusion operators, we set $g(\cdot) = \exp(0.5*(|r_1-r_2|+1))$ for the case with 10 minutes interval and $g(\cdot) = \exp(|r_1-r_2|+1)$ for that with 30 minutes interval. Setting $k=10$, we compare the joint spatial and temporal recovery with the independent spatial recovery in Fig.~\ref{fig:temprec}.

    Note that though each snapshot $\mathbf{u}^r$ is spatially and irregularly distributed, we deliberately sort them according to their indices in each snapshot. As a result, the data appear to exhibit certain periodicity, which indeed indicates the existence of temporal correlation. It is evident from Fig.~\ref{fig:temprec} that, whereas the individual spatial recovery does not deliver any meaningful recovery of the sensory data, DECA's joint spatial and temporal recovery always give excellent results. This is the case even when the sensory data appear to be non-stationary, as shown in Fig.~\ref{fig:temprec}(c).

    \vspace{1ex}\noindent\emph{Remark: We cannot evaluate the energy efficiency for this case, as we do not have the access to the original network topology. However, with $n=54$ and $k=10$, CS aggregation is bounded to save energy compared with non-aggregation.}

  \subsection{Real Field: Temperature Distribution}
\begin{figure}[t]
      \begin{center}
        \parbox{.49\columnwidth}{\center \includegraphics[width=.4\columnwidth]{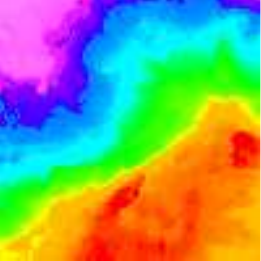}}
        \parbox{.49\columnwidth}{\center \includegraphics[width=.4\columnwidth]{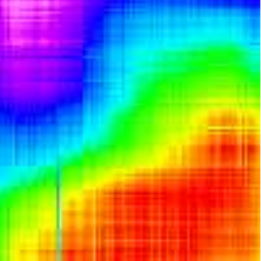}}
        \parbox{.49\columnwidth}{\center\scriptsize~(a) Original field}
        \parbox{.49\columnwidth}{\center\scriptsize~(b) Recovery field}
        \caption{Illustration of the temperature field recovery using DECA. We have $n = 2200$, four subtrees with $k_i = 160$.}
        \label{fig:weatherillu}
      \end{center}
    \end{figure}

    In this section, we validate the effectiveness of DECA over a set of temperature distribution data provided by NOAA (\url{http://www.noaa.gov/}). The NOAA datasets have been widely used by the WSN research community, e.g., \cite{Gupta-ToSN08}, \cite{LuoWSC-MobiCom09}, as they are considered as an analogy to the sensory data. The field that we take as an example is shown in Fig.~\ref{fig:weatherillu}(a), and we give one example of the final recovering result in Fig.~\ref{fig:weatherillu}(b), which accurately captures the features of the original field. Differing from the evaluations reported in Sec.~\ref{sec:peak}, in this case we also vary the WSN size by setting $\rho\in\{0.14,0.18,0.22,0.26\}$. The performance comparison with respect to different settings are plotted in Fig.~\ref{fig:weathercomp}.

    From Fig.~\ref{fig:weathercomp}(a) and (b), we can observe that, as $k$ increases, though the first level errors keep decreasing (no matter what value is taken for $\rho$),
    the second level errors become somewhat saturated. According to our experiments with matrix completion directly from the sensory data, the second level recovery errors are actually approaching such limits. Therefore, we have shown that, DECA not only enables field recovery from incomplete sensory data (due to sparse coverage of a WSN), but it also allows energy saving by performing CS aggregation in the WSN. Specifically, to recover a field represented by 10,000 samples, only 480 measurements ($<5\%$) need to be collected from the WSN.

\section{Conclusion} \label{sec:con}
  Leveraging on the recent developments in compressed sensing and harmonic analysis, we have proposed in this paper the Dual-lEvel Compressed Aggregation (DECA) framework to recover a field (of certain physical quantities) surveyed by a WSN. Although WSNs have long been deemed as powerful tools to monitor fields, DECA framework is novel because we are the first to tackle the issue of recovering fields from the aggregated version of the sensory data that are already incomplete, whereas existing proposals are mostly concerned with recovering sensory data from aggregated measurements.

  We achieve our goal by developing a novel combination of classic compressed sensing technique with matrix completion, which allows us to ``suit the medicine to the illness'' by tackling two problems with dedicated tools. Specifically, we use diffusion wavelet based compressed sensing to recover sensory data at the first level, then we apply matrix completion to recover a field at the second level. Our performance evaluations with intensive experiments have shown that DECA can achieve high recovery accuracy while still reducing energy consumption compared with traditional data collection schemes. In addition, DECA allows a WSN user to fine-tune the tradeoff between recovery accuracy and energy efficiency. Finally, by jointly exploiting spatial and temporal correlations in sensory data, DECA is applicable even to small-scale WSNs.

  We are on the way to refining the DECA framework, aiming at tuning the parameters to further improve the recovery accuracy. Also we intend to make use of the recent model-based CS to further cut down the number of measurements and hence the energy cost. Moreover, we are interested in extending DECA to 3D field and hence the 3D WSNs monitoring such fields.

\bibliographystyle{IEEEtran}
\bibliography{DECA}

\appendices
\section{Proof of Proposition~\ref{prop1}} \label{apx:prop1}

Due to the connectivity of the communication graph, combining Lemma 1.7 (iv) and (v) in~\cite{Chung97} we know that all eigenvalues of $\Lambda$ lie between 0 and 2.

The largest eigenvalue can be represented as
\begin{eqnarray*}
\sigma_\mathrm{max}(\Lambda) = \sup_f \frac{\sum_{j:(j,i)\in E}(f(j)-f(i))^2}{\sum_i f^2(i)\beta},
\end{eqnarray*}
where $f(\cdot)$ is an arbitrary real function assigned to each vertex. Therefore, $\sigma_\mathrm{max}(\Lambda)$ is a decreasing function in $\beta$. \hfill Q.E.D.

\section{Proof of Proposition~\ref{prop2}} \label{apx:prop2}

In our case, we have $n$ observed entries out of $a\times b$ samples. As suggested by (III.3) in \cite{CandesP-ProcIEEE10}, we have
\begin{eqnarray*}
\|\hat{X} - F^r\|_2 \le \left(4\sqrt{\frac{(2+q)\min(a,b)}{q}} + 2\right)\delta.
\end{eqnarray*}
Then \textit{Proposition~\ref{prop2}} follows by simply plugging $q=n/ab$, which indicates the fraction of observed entries. \hfill Q.E.D.

\section{Proof of Proposition~\ref{prop3}} \label{apx:prop3}

Here we take the results from~\cite{LeeO-ASC10}:
  \begin{quote}
  \textit{Theorem 3.4:} For a given signal $\mathbf{u}=\Psi\mathbf{w}$ with $\|\mathbf{w}\|_{\ell_0}=m$ and a clustering (permutation) scheme with parameter $\mu\in[0,1]$, the $\ell_1$ optimizer can recover $\mathbf{u}$ exactly with high probability if the number of measurements $k=\mathcal{O}(m\mu n_t\log^2n)$ where $n_t$ is the number of clusters.
  \end{quote}
The parameter $\mu$ is defined to be the maximum energy overlap between sensing matrix and sparse basis. Mathematically,
\[\mu = \max_{t,j}\sum_{i}\psi^2_{i,j}I^t_{i,j}\]
with $I^t_{i,j}=1$ indicating $\psi_{i,j}$ overlaps with cluster $t$ and otherwise $I^t_{i,j}=0$. In our case, we generate the sparse basis from diffusion wavelets, and we take the upper bound $\mu=1$. If the network is partitioned into $|\mathcal{T}|$ subtrees, we need $k=\mathcal{O}(m|\mathcal{T}|\log^2n)$ random samples to guarantee the recovery performance. \hfill Q.E.D.

\end{document}